\documentclass{iopjournal}

\usepackage{amsmath,soul,color, amssymb, cancel, amsthm, bbm}
\usepackage{wasysym, enumitem, graphicx, wrapfig, multirow, makecell, float, hyperref, hhline, booktabs}
\usepackage{subcaption, tikz}
\usepackage[numbers,square,sort&compress]{natbib}

\newcommand{\bfy}{\mathbf{y}}
\newcommand{\bfxi}{\pmb{\xi}}
\newcommand{\bfth}{\pmb{\theta}}
\newcommand{\bfW}{\mathbf{W}}
\newcommand{\bfb}{\mathbf{b}}
\newcommand{\bfD}{\mathbf{D}}
\newcommand{\bfF}{\mathbf{F}}

\graphicspath{{./figures/}}

\DeclareMathAlphabet{\mathpzc}{OT1}{pzc}{m}{it}
\newcolumntype{C}[1]{>{\centering\arraybackslash}m{#1}}

\begin{document}

\articletype{Paper}

\title{Comparative study of ensemble-based uncertainty quantification methods for neural network interatomic potentials}

\author{
    Yonatan Kurniawan$^{1,\dagger}$\orcid{0000-0002-5369-5710},
    Mingjian Wen$^2$\orcid{0000-0003-0013-575X},
    Ellad B. Tadmor$^{3,*}$\orcid{0000-0003-3311-6299},
    and
    Mark K. Transtrum$^{4}$\orcid{0000-0001-9529-9399}
}

\affil{$^1$Department of Physics and Astronomy, Brigham Young University, Provo, UT 84604, United States of America}

\affil{$^2$Institute of Fundamental and Frontier Sciences, University of Electronic Science and Technology of China, Chengdu, 611731 Sichuan, China}

\affil{$^3$Department of Aerospace Engineering and Mechanics, University of Minnesota, Minneapolis, 55455 MN, United States of America}

\affil{$^4$Cross Stream Consulting, Springville, 84663 UT, United States of America}

\affil{$^\dagger$Present address: Department of Materials Science and Engineering, University of Toronto, Toronto, ON M5S1A1, Canada}

\affil{$^*$Author to whom any correspondence should be addressed.}

\email{tadmor@umn.edu}


\begin{abstract}
    Machine learning interatomic potentials (MLIPs) enable atomistic simulations with near first-principles accuracy at substantially reduced computational cost, making them powerful tools for large-scale materials modeling.
    The accuracy of MLIPs is typically validated on a held-out dataset of \emph{ab initio} energies and atomic forces.
    However, accuracy on these small-scale properties does not guarantee reliability for emergent, system-level behavior---precisely the regime where atomistic simulations are most needed, but for which direct validation is often computationally prohibitive.
    As a practical heuristic, predictive precision---quantified as inverse uncertainty---is commonly used as a proxy for accuracy, but its reliability remains poorly understood, particularly for system-level predictions.
    In this work, we systematically assess the relationship between predictive precision and accuracy in both in-distribution (ID) and out-of-distribution (OOD) regimes, focusing on ensemble-based uncertainty quantification methods for neural network potentials, including bootstrap, dropout, random initialization, and snapshot ensembles.
    We use held-out cross-validation for ID assessment and calculate cold curve energies and phonon dispersion relations for OOD testing.
    These evaluations are performed across various carbon allotropes as representative test systems.
    We find that uncertainty estimates can behave counterintuitively in OOD settings, often plateauing or even decreasing as predictive errors grow.
    These results highlight fundamental limitations of current uncertainty quantification approaches and underscore the need for caution when using predictive precision as a stand-in for accuracy in large-scale, extrapolative applications.
\end{abstract}

\section{Introduction}
\label{sec:introduction}

Machine learning interatomic potentials (MLIPs) have emerged as powerful tools in computational materials science, offering a promising alternative to traditional approaches for simulating atomic-scale systems.
First-principles methods, such as density functional theory (DFT), provide high-fidelity predictions of material properties but are computationally prohibitive for large systems or long simulation times.
Classical empirical potentials, while computationally efficient, often lack the flexibility and transferability needed to accurately model diverse materials behavior.
MLIPs bridge this gap by leveraging machine learning algorithms to learn the underlying potential energy surface (PES) directly from quantum-mechanical reference data.
These models enable energy and force evaluations that are orders of magnitude faster than first-principles methods, while maintaining comparable levels of accuracy \cite{behler_generalized_2007_edited,thompson_spectral_2015,shapeev_moment_2016,lindsey_chimes_2017,drautz_atomic_2019,wen_uncertainty_2020,bartok_machine_2018,batzner_e3-equivariant_2022,chen_universal_2022,deng_chgnet_2023}.

Modern MLIPs are data-driven, black-box models designed to replicate the PES of atomic systems by learning directly from a quantum-mechanical dataset.
Training these models typically involves fitting to small-scale quantities, such as atomic forces, energies, and sometimes stresses, computed for a diverse set of atomic configurations \cite{ercolessi_interatomic_1994}.
The training configurations are often obtained from molecular dynamics trajectories, random structure sampling, or perturbations around equilibrium geometries.
Once trained, MLIPs are deployed to study macroscopic material properties of interest that emerge from simulations over much larger time and length scales, such as elastic properties\cite{pasianot_empirical_1991,allred_elastic_2004,tasnadi_efficient_2021}, thermal conductivity \cite{arabha_recent_2021,broido_lattice_2005,korotaev_accessing_2019}, and defect dynamics \cite{bertin_crystal_2023,page_atomistic_2024,serafin_grain_2025}.

The performance of MLIPs is often assessed by evaluating their prediction accuracy, typically quantified using error metrics that compare model predictions against ground truth DFT values, for example.
In this framework, lower error indicates higher accuracy.
Initial assessments usually focus on small-scale property predictions---energies and forces---evaluated on in-distribution (ID) samples, which consist of atomic configurations similar to those seen during training (e.g., a held-out test set).
However, high accuracy on ID samples does not necessarily imply high accuracy on out-of-distribution (OOD) samples, which are often represented by downstream, large-scale material property predictions \cite{tavazza_uncertainty_2021}.
This discrepancy stems from the mismatch in simulation scales between training and downstream application.
Large-scale property simulations frequently involve exploration into high-dimensional regions of configuration space that are poorly represented in the training set.
Moreover, such properties may depend sensitively on specific features of the PES that are difficult to sample accurately, such as saddle points corresponding to transition states \cite{samanta_sampling_2014,asgeirsson_exploring_2020} and high-energy regions in high pressure simulations \cite{cammi_studying_2022,lindsey_chimes_2025}.
Although this motivates the need to assess accuracy on both ID and OOD samples, generating such ground truth data---whether quantum-mechanical or experimental---is often prohibitively expensive or even infeasible.

In addition to accuracy, it is also important to assess model precision through estimates of predictive uncertainty, where lower uncertainty indicates higher precision.
Precision reflects the consistency of a model’s predictions under various sources of variability, such as those arising from the training data, model architecture, initialization, and optimization.
Various uncertainty quantification (UQ) methods have been developed to capture these effects in MLIPs.
For example, techniques like mean-variance estimation \cite{tan_single-model_2023}, Gaussian mixture models \cite{zhu_fast_2023}, and quantile regression \cite{bilbrey_uncertainty_2025} aim to estimate aleatoric uncertainty, i.e., irreducible uncertainty from inherent variability in the data.
These methods learn such variability directly from the training data, including implicit sources of variability arising from choices of DFT exchange-correlation functionals and numerical tolerances \cite{henkel_uncertainty_2021, dai_uncertainty_2024}.
In contrast, ensemble-based approaches \cite{li_uncertainty_2019, wen_uncertainty_2020} and Bayesian neural networks \cite{tran_methods_2020} target epistemic uncertainty, which arises from limited data coverage, model misspecification, and parameter uncertainty.
Ensemble methods, in particular, are popular due to their simplicity, model-agnostic implementation, and practical effectiveness.
Additionally, distance-based metrics \cite{vita_ltau-ff_2024, hu_robust_2022} and Gaussian process models are often employed to detect OOD configurations and estimate model confidence in previously unexplored regions of configuration space.

\begin{figure}[!hbt]
    \centering
    \includegraphics[width=0.4\textwidth]{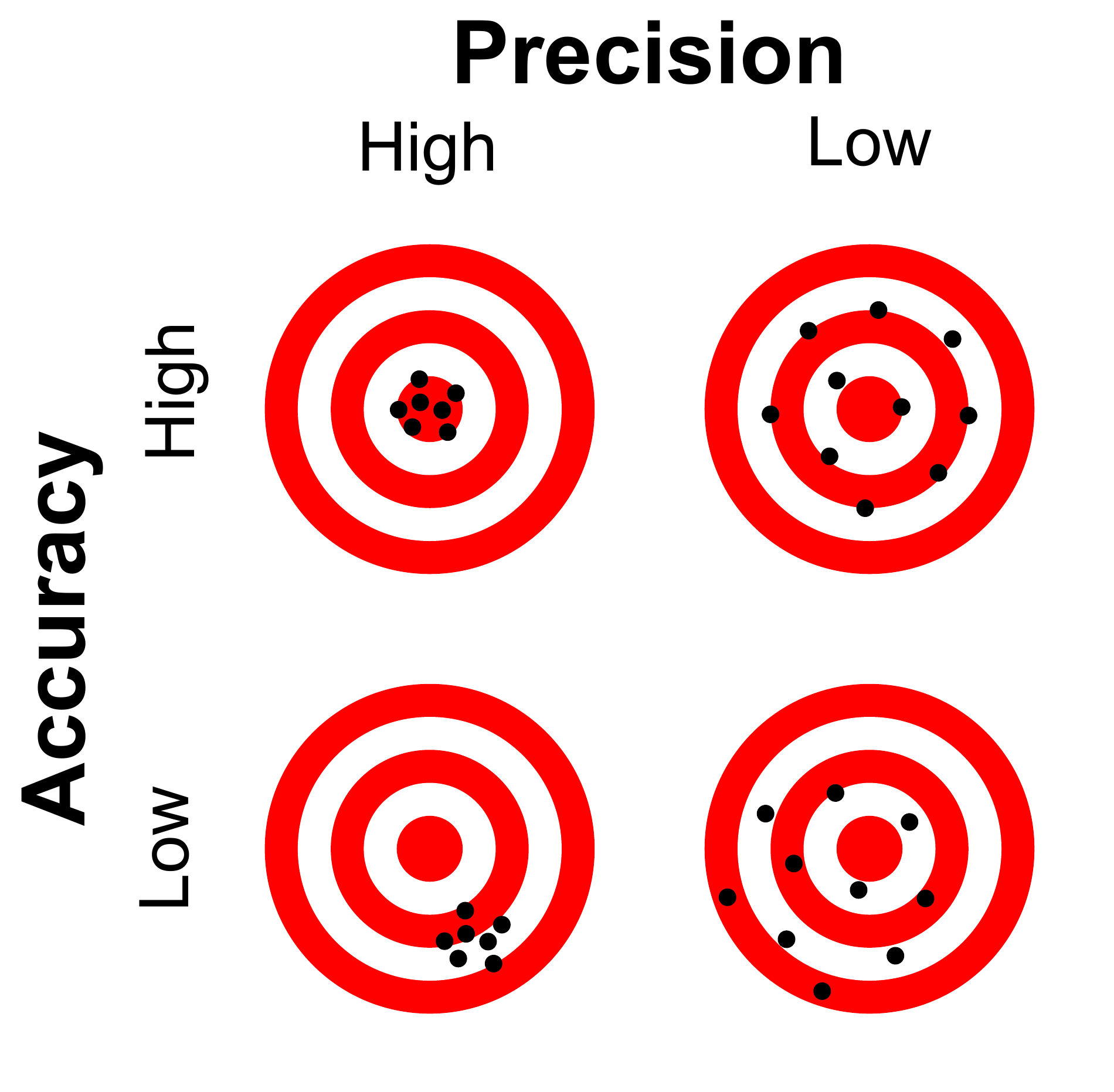}
    \caption[Illustration of precision and accuracy using a dartboard metaphor]{
	Illustration of precision and accuracy using a dartboard metaphor.
	Each panel shows a different combination of prediction accuracy (closeness to the true value, indicated by the bullseye) and precision (spread of predictions).
	This illustration emphasizes that high precision does not necessarily imply high accuracy.
    }
    \label{fig:precision_accuracy}
\end{figure}

The use of prediction precision as a proxy for accuracy has been suggested as a practical solution to the difficulty of evaluating accuracy, especially in the context of large-scale material property predictions \cite{frederiksen_bayesian_2004}.
Although estimating prediction precision (i.e., uncertainty) can be computationally intensive---often requiring multiple model evaluations---it remains far less costly than quantum-mechanical or experimental validation.
However, precision and accuracy are fundamentally distinct concepts and do not necessarily correlate (see Fig.~\ref{fig:precision_accuracy}).
While one might expect accurate predictions to be accompanied by high precision (top-left panel) and inaccurate predictions by low precision (bottom-right panel), this relationship does not always hold.
For example, a model may produce accurate material property predictions on average, yet still exhibit a large ensemble spread (top-right panel).
Conversely, a prediction can appear highly precise, with low ensemble variance, yet deviate significantly from the true value, resulting in \emph{overconfident} predictions (bottom-left panel).
Recognizing this potential disconnect is crucial for evaluating when uncertainty estimates can be interpreted meaningfully and when they might give a false sense of confidence or caution.

Several studies have evaluated the reliability of UQ methods for MLIPs.
Kahle and Zipoli~\cite{kahle_quality_2022} examined whether ensemble uncertainties reflect prediction errors for OOD configurations and showed that their reliability can depend strongly on the atomic system and model architecture.
Carrete et al.~\cite{carrete_deep_2023} compared several ensemble constructions in OOD molecular-dynamics and active-learning applications, finding that more elaborate uncertainty estimators did not consistently outperform simpler committees in identifying difficult configurations.
Tan et al.~\cite{tan_single-model_2023} compared model ensembles with several single-model UQ approaches and found that no method performed consistently best across all settings.
More recently, Wimer et al.~\cite{wimer_benchmarking_2026} benchmarked neural network interatomic potentials targeting epistemic, aleatoric, and combined uncertainty using prediction-accuracy and uncertainty-calibration metrics.
These studies demonstrate that UQ performance depends on the data regime, evaluation criterion, model configuration, and intended application.

In this paper, we investigate when prediction precision can serve as a reliable surrogate for prediction accuracy, particularly in the OOD domain.
We conduct a comparative study of several ensemble-based UQ methods for MLIPs, highlighting key considerations when using precision as a proxy for accuracy.
To this end, our study involves developing MLIPs, generating multiple MLIP ensembles, and applying them within widely used atomistic simulation software to compute large-scale material properties.
These tasks are streamlined by the infrastructure within the Open Knowledgebase of Interatomic Models (OpenKIM) project, which enables seamless integration of MLIPs into simulation workflows \cite{Tadmor_Elliott_Sethna_Miller_Becker_2011,elliott:tadmor:2011}.
Furthermore, this study aligns with OpenKIM’s broader goal of promoting the reliable, accessible, and reproducible development and evaluation of interatomic potentials.

The remainder of the paper is organized as follows.
Section~\ref{sec:methods} introduces the MLIP architecture used in this study, with a particular focus on the neural network interatomic potential (NNIP).
We then describe the ensemble-based UQ methods employed and the set of material properties used to evaluate uncertainty estimates in both ID and OOD domains.
Considering both domains is crucial for gaining a comprehensive understanding of the relationship between precision and accuracy.
Section~\ref{sec:results} presents our findings on the uncertainty behavior of each ensemble method across the selected properties.
Section~\ref{sec:discussion} analyzes the observed trends and provides possible explanations in terms of model extrapolation, along with a discussion of caveats regarding uncertainty estimation in extrapolation regimes.
The conclusion of this work is given in Sec.~\ref{sec:conclusion}.
Finally, we emphasize that our goal is not to propose methods for mitigating issues regarding uncertainty, but rather to characterize and analyze the behavior of ensemble-based UQ methods in this context.


\section{Methods}
\label{sec:methods}

In this section, we describe the methods used to evaluate the relationship between prediction precision and accuracy in MLIPs.
We begin by introducing the NNIP, which serves as the MLIP architecture in this study.
This architecture offers a flexible functional form for approximating complex, high-dimensional PESs and is widely used due to its relative simplicity and effectiveness \cite{behler_perspective_2016, wang_machine_2024}.
Next, we describe the training process, including the dataset, loss function, and optimization strategy.
We then introduce ensemble-based UQ and describe the specific ensemble methods compared in this work.
Finally, we present the set of material properties---both small- and large-scale quantities---used to evaluate the uncertainty estimates produced by the ensemble models.

The workflows described above are supported by infrastructure developed through the OpenKIM project.
The OpenKIM project aims to promote reliability, accessibility, and reproducibility in atomistic simulations by providing standardized interfaces, a curated repository of interatomic potentials, and comprehensive testing tools for interatomic models \cite{Tadmor_Elliott_Sethna_Miller_Becker_2011}.
The KIM Application Programming Interface (KIM-API) \cite{elliott:tadmor:2011} enables seamless integration of KIM-compliant potentials with widely used atomistic simulation packages, such as ASE \cite{larsen_atomic_2017} and LAMMPS \cite{LAMMPS}.
The OpenKIM repository hosts an expanding collection of interatomic potentials, including both empirical models and MLIPs.
Submitted models are automatically validated through a verification pipeline that ensures compliance with interface standards and tests for essential physical constraints.

The NNIPs used in this study are developed using the KIM-based Learning-Integrated Fitting Framework (KLIFF), a Python package developed under the OpenKIM project for training empirical potentials and MLIPs \cite{wen_kliff_2022}.
Potentials trained with KLIFF are KIM-compliant, allowing direct integration with any simulation code that supports the KIM API.
It also provides built-in support for various UQ methods for both empirical models \cite{kurniawan_extending_2022} and MLIPs \cite{wen_uncertainty_2020}, facilitating systematic UQ studies for interatomic potentials.
All UQ ensembles in this work are generated using various functionalities in KLIFF.

\subsection{Neural network interatomic potential}
\label{sec:nnip}

\begin{figure*}[!hbt]
    \centering
    \includegraphics[width=0.8\textwidth]{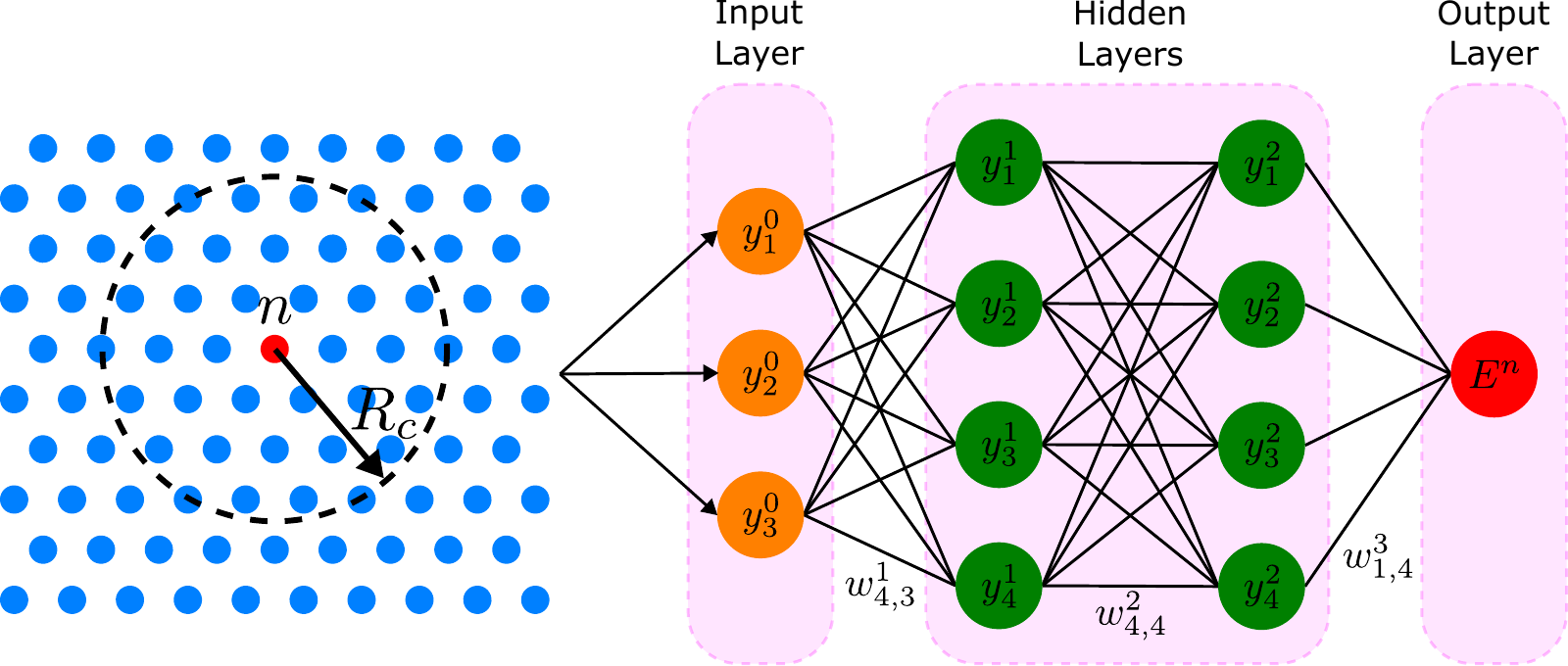}
    \caption[Graph representation of a neural network interatomic potential]{
	Graph representation of a neural network interatomic potential.
	$y^l_\alpha$ is the $\alpha$-th element of $\bfy^l$ and $w^l_{\alpha, \beta}$ is the element of $\bfW^l$ on $\alpha$-th row and $\beta$-th column.
    }
    \label{fig:nnip_cartoon}
\end{figure*}

The total energy of a configuration containing $N$ atoms is modeled as the sum of atomic energy contributions,
\begin{equation}
    \label{eq:config_energy}
    E = \sum_{n=1}^N E_n (\bfxi^n),
\end{equation}
where each atomic contribution $E_n$ is a function of the local atomic environment of the respective atom, represented by a descriptor vector $\bfxi^n$ (more details about the atomic descriptor are given in the supplementary material).
These atomic energies are approximated using a neural network (NN) model.
Figure~\ref{fig:nnip_cartoon} illustrates a schematic of a commonly used NN architecture, the multilayer perceptron, in which each node in layer $l$ is fully connected to every node in the preceding layer $(l-1)$.
The activations $\bfy^l$ of layer $l$, is computed as
\begin{equation}
    \label{eq:nn_mapping}
    \bfy^l = \sigma_l \left( \bfW^l \bfy^{(l-1)} + \bfb^l \right),
\end{equation}
where $\bfW^l$ is the weight matrix connecting layers $l-1$ and $l$, $\bfb^l$ is the corresponding bias vector, and $\sigma_l(\cdot)$ is a nonlinear activation function applied element-wise.
As previously mentioned, the input to the NNIP is the descriptor vector $\bfxi^n$, and the output is the predicted energy contribution $E_n$ of atom $n$.
The force acting on atom $n$ is calculated as the negative gradient of Eq.~(\ref{eq:config_energy}) with respect to the atom's Cartesian coordinates.

The NNIP architecture used in this study consists of three hidden layers, each with 128 nodes, and employs the hyperbolic tangent activation function to ensure smooth and differentiable outputs.
Local atomic environments encoded using atom-centered symmetry functions \cite{behler_atom-centered_2011_edited} serve as inputs to the neural network model.
Dropout with a rate of $0.1$ is applied to all hidden layers as a regularization technique during training.
In addition, dropout is also employed as one of the uncertainty quantification methods considered in this work, where it is used to generate an ensemble of predictions.
Details on the implementation of dropout for both regularization and uncertainty estimation are provided in Sec.~\ref{sec:dropout}.

\subsection{Potential training}
\label{sec:training}

The NNIP is trained on a carbon dataset used in a previous study \cite{wen_uncertainty_2020,Wen2020_dataset}.
The dataset comprises atomic configurations from various carbon allotropes, including monolayer and bilayer graphene, graphite, and diamond (see Table~\ref{tab:dataset}).
These configurations were sampled from \emph{ab initio} molecular dynamics trajectories at temperatures ranging from $300$ to $900~\mathrm{K}$, with additional trajectories at $1{,}500~\mathrm{K}$ for graphite and monolayer graphene.
The dataset also includes uniformly biaxially strained monolayer graphene configurations, with strain values ranging from $-3\%$ to $+3\%$, together with corresponding configurations generated by perturbing the atomic positions.
For each configuration, the reference total energy and atomic forces were computed using DFT with the Perdew--Burke--Ernzerhof (PBE) exchange--correlation functional \cite{perdew_generalized_1996}.
For bilayer graphene and graphite, the many-body-dispersion correction was included to account for interlayer van der Waals interactions.
The calculations used a plane-wave energy cutoff of $500~\mathrm{eV}$, and periodic slab cells for monolayer and bilayer graphene included approximately $30~\text{\AA}$ of vacuum perpendicular to the layers.
Reciprocal space was sampled using $\Gamma$-centered Monkhorst--Pack meshes selected to ensure convergence of the total energy.

\begin{table*}[!hbt]
    \centering
    \caption[Number of configurations in the carbon dataset]{Number of configurations in the carbon dataset.}
    \label{tab:dataset}
    \begin{tabular}{l @{\hspace{1cm}} r @{\hspace{0.2cm}} r @{\hspace{0.2cm}} r}
      \toprule
      \multirow{2}{*}{Structure} & \multicolumn{2}{c}{Number of configurations} & \multirow{2}{*}{\hspace*{0.5cm}\strut\vspace{2ex} Number of atoms} \\
      \cline{2-3}
				 & Training set & Test set & per configuration \\
      \midrule
      Diamond & 759  & 84 & 64 \\
      Graphite & 661 & 81 & 72 \\
      Monolayer graphene & 2,181 & 185 & 2-32 \\
      Bilayer graphene & 743 & 94 & 52-76 \\
      \bottomrule
    \end{tabular}
\end{table*}

The potential parameters $\bfth$, i.e., the weights and biases of the NNIP, are optimized by minimizing a loss function
\begin{equation}
    \label{eq:loss_function}
    \mathcal{L}(\bfth) = \frac{1}{2} \sum_{m=1}^M \left[ \left(r_m^E(\bfth)\right)^2 + \Vert r_m^F(\bfth) \Vert_2^2  \right],
\end{equation}
where $r_m^E$ and $r_m^F$ denote the residuals, i.e., weighted error, of the configuration energy and atomic forces, respectively.
The energy residual is defined as
\begin{equation}
    \label{eq:energy_residual}
    r_m^E(\bfth) = \frac{1}{N_m \sigma_E} \left( E_m^{\text{DFT}} - E_m(\bfth) \right),
\end{equation}
where $N_m$ is the number of atoms in configuration $m$.
The factor of $1/N_m$ ensures that each configuration contributes equally in the loss function, regardless of its size, and the denominator factor $\sigma_E = 1~\text{eV}$ ensures that the energy residual is dimensionless.
The contribution from atomic forces is given by the squared $\ell_2$-norm of the force residual vector,
\begin{equation}
    \label{eq:forces_residual}
    \Vert r_m^F(\bfth) \Vert_2^2 = \sum_{n=1}^{N_m} \sum_{i=1}^3 \left( \left( r_m^{\bfF}(\bfth) \right)_{3(n-1)+i} \right)^2,
\end{equation}
with each element of the residual vector defined as
\begin{equation}
    \label{eq:vector_forces_residual}
    \begin{aligned}
	\left( r_m^F(\bfth) \right)_{3(n-1)+i} = \frac{1}{N_m \sigma_F} \left( (F_i^n)_m^{\text{DFT}}- (F_i^n)_m(\bfth) \right),
    \end{aligned}
\end{equation}
where the subscript $i$ indexes the Cartesian component of the force vector.
An additional factor of $\sigma_F = \sqrt{10}~\text{eV/\AA}$ is included in the denominator to approximately balance the contributions of the energy and force term in the loss function, and to ensures that the force residual term is also dimensionless.

The dataset comprises a total of 4{,}788 atomic configurations.
A randomly selected 90\% of the data is used to train the model by minimizing the loss in Eq.~\ref{eq:loss_function} with the Adam optimizer~\cite{kingma_adam_2017}.
We employ a batch size of 100; the learning rate is set to $10^{-3}$ for the first 5{,}000 epochs and reduced to $10^{-4}$ for the remaining 35{,}000 epochs.
For most ensemble methods, the final model is selected based on the epoch with the lowest loss on the test set (which consists of the remaining 10\% of the data).
We also note that, to ensure a fair comparison of the various UQ methods, we fix the model architecture as described in Section~\ref{sec:nnip}, eliminating the need for a validation set for hyperparameter selection, unlike the typical machine learning pipeline that uses separate training, validation, and test sets.

\subsection{Ensemble-based uncertainty quantification methods}
\label{sec:uq_methods}

Ensemble-based methods are among the most widely used types of UQ methods for NNIPs.
The key idea is to construct an ensemble of statistically similar models by introducing some form of variation during training.
The model prediction and its associated uncertainty are estimated using the mean and standard deviation of the ensemble predictions,
\begin{align}
  \label{eq:mean_std_predictions}
  \mu_y &= \frac{1}{S} \sum_{s=1}^S y_s \\
  \sigma_y^2 &= \frac{1}{S-1} \sum_{s=1}^S (y_s - \mu_y)^2,
\end{align}
where $y_s$ denotes the prediction from the $s$-th ensemble member and $S$ is the total number of models in the ensemble.
Several methods have been developed to evaluate and calibrate uncertainties \cite{kuleshov_accurate_2018,levi_evaluating_2020}.
In this work, however, $\sigma_y$ is used directly as the raw ensemble spread, without post-hoc calibration to allow us to compare the uncertainty behavior generated directly by the different ensemble-construction methods without introducing an additional dependence on the calibration procedure and calibration dataset.

There are various sources of randomness involved in the training of NNIPs, and different ensemble-based UQ methods leverage different aspects of this randomness.
For example, the bootstrap ensemble captures variability in the training data by resampling the dataset.
The Monte Carlo dropout ensemble reflects variability in the model architecture and the function space it can represent.
The random initialization ensemble captures sensitivity to the initial model parameters.
The snapshot ensemble accounts for the stochasticity of the training process, particularly that introduced by stochastic gradient descent (SGD).

\begin{figure*}[!hbt]
    \centering
    \includegraphics[width=0.95\textwidth]{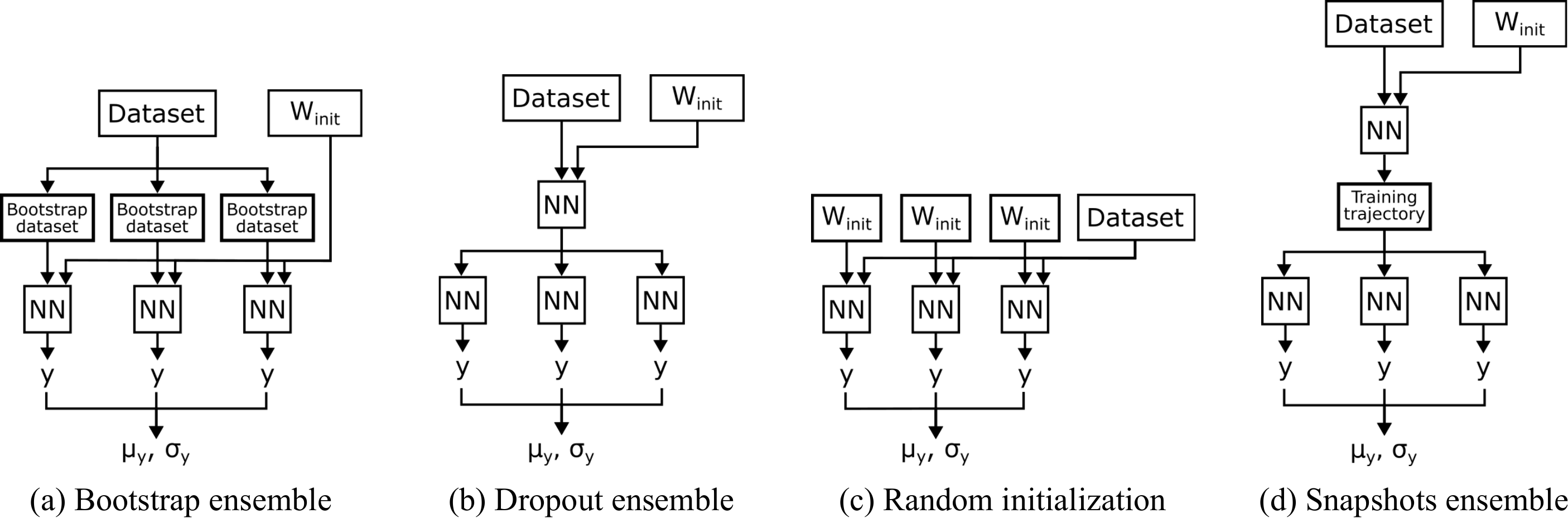}
    \caption[Illustration of ensemble-based UQ methods for NNIPs]{
	Illustration of ensemble-based UQ methods for NNIPs compared in this work.
	(a) In bootstrap ensemble, multiple bootstrap datasets are generated by sampling the original dataset with replacement, and separate NNIPs are trained on each dataset.
	(b) With dropout ensemble, different dropout masks are applied during prediction, effectively deactivating different subsets of nodes in the network to form an ensemble.
	(c) In the random initialization ensemble, multiple NNIPs are independently initialized with different weights and biases, and then trained on the same dataset.
	(d) Finally, the snapshot ensemble is generated by saving model instances (snapshots) at different epochs along the training trajectory.
    }
    \label{fig:ensemble_methods}
\end{figure*}

In this work, we compare the aforementioned UQ methods to study when prediction precision can be reliably used as a proxy for accuracy.
Following Wen and Tadmor \cite{wen_uncertainty_2020}, we generate 100 models per ensemble to ensure that the predictive mean and uncertainty estimates converge within an acceptable tolerance.
Figure~\ref{fig:ensemble_methods} schematically illustrates each approach.
Comparing these methods offers a more comprehensive understanding of when and to what extent the prediction accuracy correlates with model precision.

\subsubsection{Bootstrap}
\label{sec:bootstrap}

A commonly used method to introduce variability into the training process is bootstrapping, which involves generating synthetic datasets by resampling the original dataset with replacement \cite{efron_bootstrap_1979}.
Each bootstrap sample serves as the training data for a separate model in the ensemble, creating diverse training conditions that help prevent models from becoming overly dependent on specific data points.
In the context of neural networks, where the loss landscape is highly non-convex, it is important to maintain consistent training settings across all models to isolate the source of variability in the predictions.

A fundamental assumption underlying bootstrapping is that the data are independently and identically distributed \cite{efron_bootstrap_1979}.
However, when training interatomic potentials, the dataset typically comprises many atomic configurations, each containing multiple labeled data points, e.g., the force vector components on individual atom.
Due to interatomic interactions, the atomic forces within a single configuration are inherently correlated.
To mitigate this issue, bootstrapping is performed at the configuration level rather than on individual data points \cite{apeterson_addressing_2017}.
Nevertheless, this strategy does not fully eliminate correlation-related biases if the configurations themselves are not independent, for instance, when they are sequential snapshots from a single molecular dynamics trajectory.
In such cases, the resulting ensemble predictions may exhibit overconfidence.

As part of this study, we have added support for bootstrap UQ in KLIFF to facilitate UQ studies for MLIPs.
Our implementation follows the recommendations outlined above, including using consistent optimizer settings across all ensemble trainings and employing configuration-level resampling as the default strategy.
Alternative sampling strategies can be readily defined to accommodate specific applications.

\subsubsection{Monte Carlo Dropout}
\label{sec:dropout}

Dropout was originally introduced as a regularization technique to prevent NN models from overfitting by randomly deactivating (or ``dropping out'') nodes during each training epoch with a fixed probability, i.e., dropout ratio, of $p$ \cite{srivastava_dropout_2014}.
This process prevents individual nodes from becoming overly dominant, i.e., having disproportionately large weights, thereby encouraging the network to learn more generalized features.
Dropout is implemented in layer $l$ of a NN model by defining a diagonal dropout mask matrix $\mathbf{D}^l$, where each diagonal element is a random binary variable (0 or 1).
Each element is set to zero with probability $p$.
The activations of layer $l$ are then computed as (replacing Eq.~(\ref{eq:nn_mapping}))
\begin{equation}
    \label{eq:dropout_layer}
    \bfy^l = \sigma_l \left( \bfW^l (\bfD^l \bfy^{(l-1)}) + \bfb^l \right),
\end{equation}
where the dropout matrix $\bfD^l$ is applied to the input vector $y^{(l-1)}$, effectively deactivating a subset of input nodes to layer $l$.

Dropout ensemble extends this regularization technique by employing it to generate an ensemble of neural network models \cite{gal_dropout_2016}.
Each ensemble member is assigned a unique set of dropout masks, resulting in different subsets of nodes being deactivated across the models.
As the dropout rate increases, the models in the ensemble become more diverse, introducing greater variability in their architectures and higher uncertainty in their predictions.
However, the model is also forced to rely on a smaller subset of features, which may degrade performance if too many nodes are dropped.

\subsubsection{Random initialization}
\label{sec:randinit}

The loss surface of a NNIP is highly nonconvex, causing distinct optimization runs to converge to different parameter values that yield similar loss \cite{soudry_exponentially_2017,kwak_understanding_2018}.
Furthermore, this redundancy reflects the overparameterized nature of modern machine learning architectures, where the same training data can be fit in many distinct ways.
Although models converging to different minima may achieve comparable accuracy on training and validation data, they often diverge significantly on OOD inputs or edge cases, which undermines their generalizability.

A straightforward strategy to capture this variability is to train an ensemble of networks with identical architectures and hyperparameters but different random initial weights and biases \cite{lakshminarayanan_simple_2017}.
Such ensembles integrate seamlessly into any standard training pipeline and can be applied to any modern machine learning model.
Their primary drawback, similar to bootstrap-based methods, is computational cost since each model in the ensemble must be trained independently.
Nevertheless, random initialization ensemble has been shown to consistently deliver significant gains in predictive performance and uncertainty calibration, rivaling those of Bayesian NNs \cite{lakshminarayanan_simple_2017}.

\subsubsection{Snapshot}
\label{sec:snapshot}

The snapshot ensemble method leverages the inherent stochasticity introduced by mini-batch sampling in SGD during training.
Chaudhari and Soatto  \cite{chaudhari_stochastic_2018} show that the optimization trajectory of SGD exhibits behavior analogous to Bayesian sampling, where the process effectively samples from a posterior distribution over model parameters.
They further demonstrate that the level of noise in SGD is proportional to the ratio between the learning rate and the mini-batch size.
Building on this idea, we construct an ensemble by capturing model snapshots at various points along the training trajectory \cite{huang_snapshot_2017,izmailov_averaging_2019,maddox_simple_2019}.

Similar to Bayesian sampling, several considerations must be taken into account when generating a snapshot ensemble.
First, we aim to reduce the influence of the initial training conditions and avoid including suboptimal models in the ensemble.
Thus, we discard snapshots from the early training epochs, akin to a burn-in period, and begin capturing models only after the training loss has plateaued.
However, some variations of the method incorporate early-stage snapshots to capture additional sources of variation or learning dynamics \cite{vita_ltau-ff_2024}.
To ensure the collected snapshots are approximately independent, we save them at regular, large intervals, specifically every 100 epochs.
This approach allows us to efficiently sample a diverse set of high-performing models from a single training session.

\subsection{Target properties}
\label{sec:target-properties}

We evaluate the performance of the UQ methods by comparing their predictions and associated uncertainties on both ID and OOD samples.
For the former, we assess configuration energies and atomic forces on the training and test sets as a form of validation.
For the latter, we examine predictions and uncertainties for large-scale material properties.
These large-scale properties emerge from many-atom interactions and typically involve configurations unseen during training.
By analyzing both ID and OOD, we obtain a more comprehensive view of each method’s performance, as discussed in Sec.~\ref{sec:discussion}.

\begin{figure*}[!hbt]
    \centering
    \includegraphics[width=\textwidth]{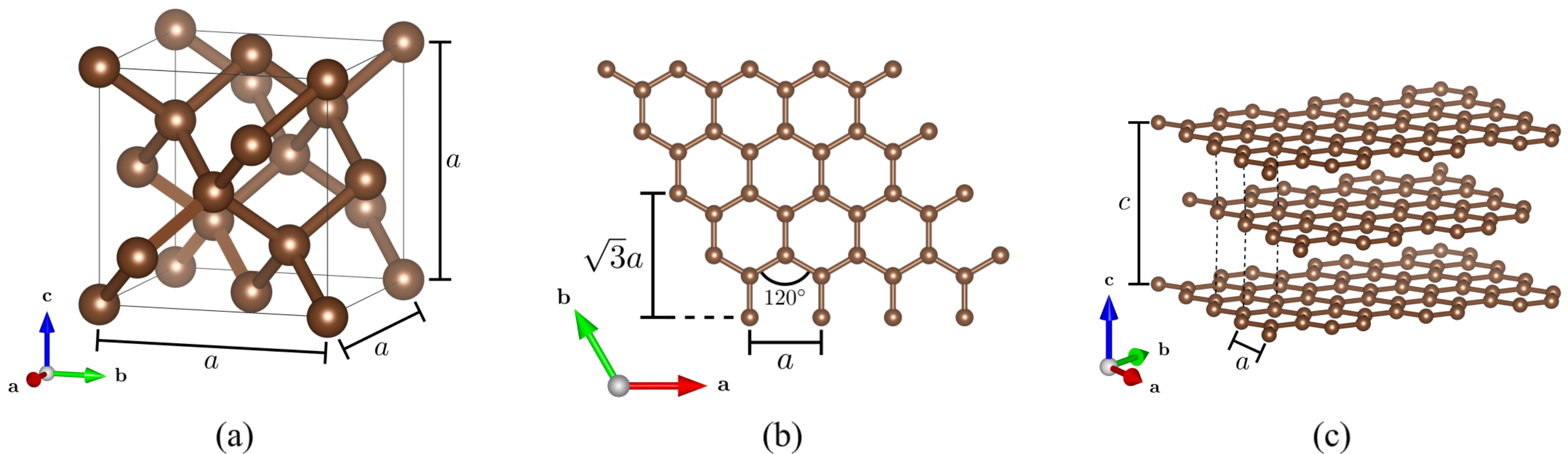}
    \caption[Illustration of several crystal structures of carbon]{
	Illustration of several crystal structures of carbon: (a) diamond, (b) monolayer graphene, and (c) graphite.
	In the diamond lattice, each carbon atom is tetrahedrally bonded to four other carbon atoms, forming a three-dimensional cubic structure with lattice parameter denoted by $a$.
	A single graphene sheet consists of carbon atoms arranged in a two-dimensional honeycomb lattice, with an in-plane lattice parameter denoted by $a$ and an angle of $120^\circ$ between carbon-carbon bonds.
	Graphite is composed of stacked layers of graphene, with the layers arranged in a staggered ABAB pattern.
	The in-plane lattice parameter of graphite is denoted by $a$, while the interlayer spacing between adjacent graphene layers is denoted as $c/2$, with the layers held together by weak van der Waals forces.
    }
    \label{fig:structures}
\end{figure*}

As representative large-scale properties, we compute the energy cold curve and the phonon dispersion relation for three distinct carbon allotropes: diamond, monolayer graphene, and graphite (see Fig.~\ref{fig:structures} for illustrations of each structure).
These properties are computationally inexpensive and functionally distinct from the training data, yet provide valuable insight into the structural, elastic, and vibrational characteristics of the materials.
Details on how each property is computed are provided below.

\subsubsection{Energy cold curve}
\label{sec:coldcurve}

The energy cold curve describes the system's energy as a function of the lattice parameter at 0~K, offering insight into the material’s structural properties.
The minimum of this curve corresponds to the equilibrium lattice constant(s), and the associated energy per atom, i.e., the cohesive energy, quantifies the energy required to disassemble the crystal into isolated atoms.
Furthermore, the curvature near the equilibrium point reflects the material’s resistance to deformation, providing information about its elastic properties.

The energy cold curve is computed by varying the lattice parameter $a$ and evaluating the energy per atom for each configuration.
Unless otherwise specified, $a$ is perturbed by $\pm 10\%$ from the equilibrium value reported in the Materials Project repository \cite{10.1063/1.4812323}.
For the diamond structure, cubic symmetry is preserved by applying isotropic strain, effectively varying $a$ uniformly in all directions.
In the case of graphene and graphite, the bond angles are fixed at $120^\circ$  to maintain the honeycomb structure, and strain is applied uniformly in the in-plane directions only.
Additionally, for graphite, we allow relaxation of the lattice along the out-of-plane direction (i.e., the $c$ parameter) to account for the weak van der Waals interactions between layers.
However, in our NNIP-based calculations, the relaxed value of $c$ exhibits negligible variation across the range of $a$.
Therefore, we only report the energy cold curve for graphite as a function of $a$ in the next section.

\subsubsection{Phonon dispersion}
\label{sec:phonon}

The other material property we examine is the phonon dispersion relation, which provides insight into the vibrational properties of the material.
Phonon dispersion curves show how the frequencies of lattice vibrations (phonons) vary with the wavevector across the Brillouin zone.
Phonons play a central role in determining key dynamical properties, such as heat capacity and thermal conductivity.

Computing the phonon dispersion curves involves calculating the force constant matrix, i.e., the Hessian of the potential energy with respect to atomic displacements.
The force constant matrix is then transformed into reciprocal space via a Fourier transform to construct the dynamical matrix.
The phonon frequencies at a given wavevector are obtained by solving the eigenvalue problem of the dynamical matrix, where the squared frequencies correspond to the eigenvalues.
In this work, we use a finite difference approach implemented in the Atomic Simulation Environment (ASE) Python package to perform these calculations \cite{alfe_phon_2009, larsen_atomic_2017}.
To keep the main text concise, the resulting phonon dispersion results are presented in the supplementary material.


\section{Results}
\label{sec:results}

\subsection{In-distribution comparison}
\label{sec:in-distribution}

\begin{figure}[!hbt]
    \centering
    \includegraphics[width=0.45\textwidth]{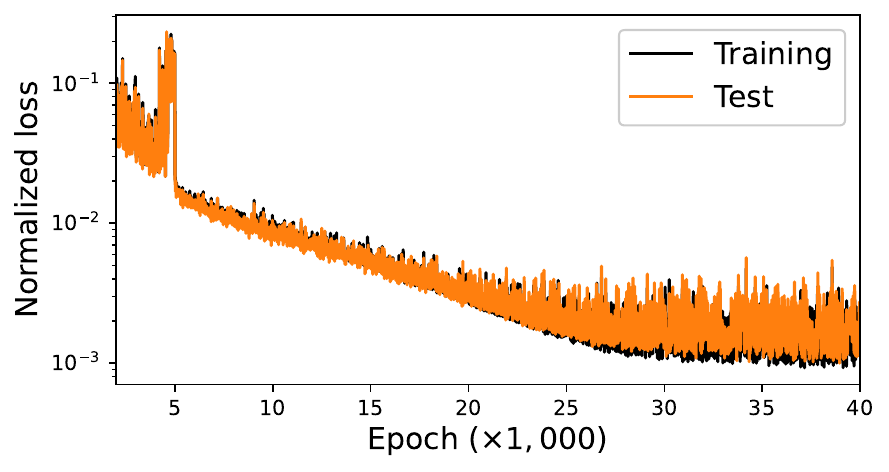}
    \caption[Loss trajectory over the training and test sets]{
        Loss trajectory over the training and test sets for a single NNIP used to generate the dropout and snapshot ensembles.
        The loss values are normalized by the number of configurations in each set.
        The comparable magnitudes of the training and test losses indicate limited overfitting.
    }
    \label{fig:loss}
\end{figure}

We first present the training performance of the ensemble models investigated in this study.
Figure~\ref{fig:loss} shows the training and test loss trajectories for the single NNIP used to generate the dropout and snapshot ensembles, which are representative of the overall training behavior.
Table~\ref{tab:rmse_mae} reports the root-mean-square error (RMSE) for energies and forces evaluated on both the training and test sets.
The close agreement between training and test losses, as well as the similar error values on both datasets, indicates that the models do not exhibit substantial overfitting and exhibit consistent performance within the ID region.
Interestingly, the bootstrap ensemble exhibits the largest energy RMSE while simultaneously yielding the smallest force RMSE.
Although the origin of this behavior is not investigated further in the present work, the resulting energy and force errors remain within the expected accuracy of DFT and are comparable across ensemble approaches.

\begin{table*}[!hbt]
    \centering
    \caption[Energy and forces RMSE on the training and test sets for each UQ ensemble model]{
	Energy and forces RMSE evaluated on the training and test sets for each UQ ensemble model.
    }
    \label{tab:rmse_mae}
    \begin{tabular}{ m{10em} C{6em} C{6em} C{6em} C{6em} }
      \toprule
      & \multicolumn{2}{c}{\textbf{Energy} (meV/atom)} & \multicolumn{2}{c}{\textbf{Forces} (meV/\AA)} \\
      \cmidrule(lr){2-5}
      & Training & Test & Training & Test \\
      \midrule
      Bootstrap             & 8.905 & 9.204 & 3.001 & 3.350 \\
      Dropout               & 6.087 & 6.252 & 5.249 & 5.559 \\
      Random initialization & 5.910 & 6.164 & 4.631 & 4.948 \\
      Snapshots             & 7.105 & 7.235 & 5.218 & 5.501 \\
      \bottomrule
    \end{tabular}
\end{table*}

\begin{figure*}[!hbt]
    \centering
    \includegraphics[width=0.9\textwidth]{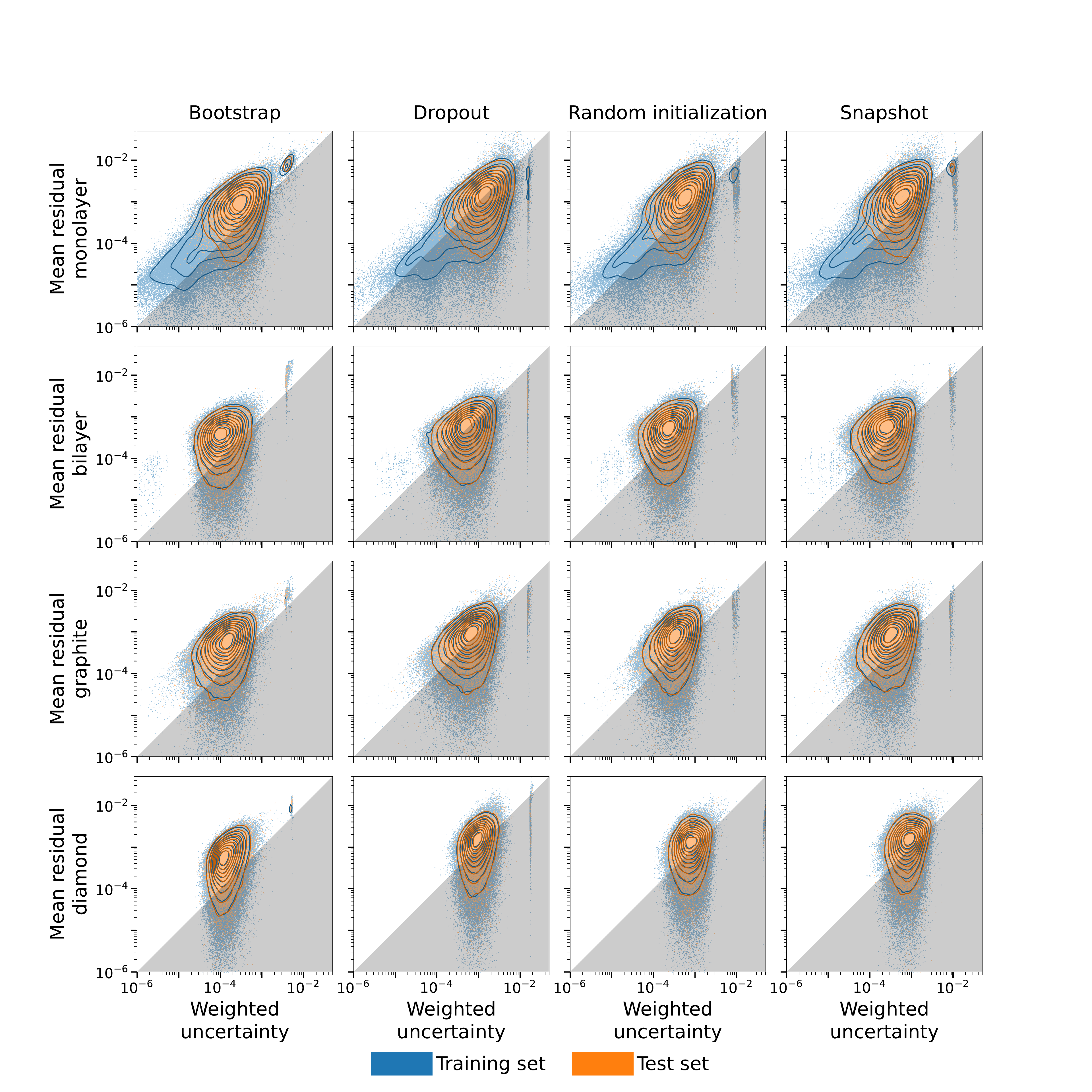}
    \caption[Parity plot comparing the residual and the uncertainty]{
        Parity plot comparing the residual (weighted absolute error) and the uncertainty (one standard deviation).
        Individual samples from the training (blue) and test (orange) sets are shown as semi-transparent markers, with kernel density estimates overlaid as contours in the corresponding colors to highlight the high-density regions.
        The columns represent different ensemble models, while the rows represent the crystal structures in the dataset.
        The grey region represents the underconfident region, where the uncertainty is larger than the residual.
        In an ideal condition, the sample distribution would lie around the diagonal, where the residual and uncertainty are correlated and approximately equal to each other.
    }
    \label{fig:accuracy_energy_forces}
\end{figure*}

Next, we evaluate and compare the accuracy and precision of these ensemble models in the ID domain.
Figure~\ref{fig:accuracy_energy_forces} shows the absolute residual vs. the uncertainty---chosen to be one standard deviation---for each data point in the training (blue) and test (orange) sets.
The individual observations are shown as semi-transparent markers to reveal tail samples.
To better represent the data distribution in the high-density regions, we additionally overlay kernel density estimates as contours in the corresponding colors.
We separate the structure types in the dataset as the rows in this figure, and compare different ensemble models in columns.
Additionally, we overlay a grey region on each plot to highlight where the uncertainty exceeds the residual.
Changing the definition of uncertainty (e.g., as two standard deviations instead) would shift the diagonal boundary vertically.

\begin{figure*}[!hbt]
    \centering
    \begin{subfigure}[b]{0.32\textwidth}
        \centering
        \includegraphics[height=\textwidth]{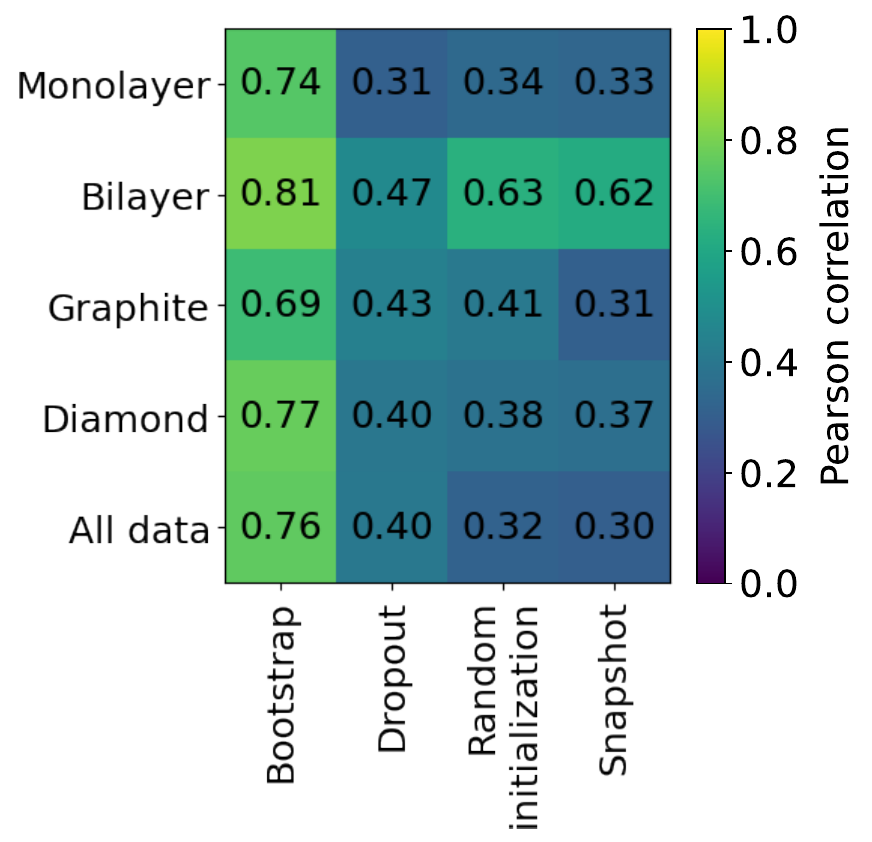}
        \caption{}
	\label{fig:correlation}
    \end{subfigure}
    \hfill
    \begin{subfigure}[b]{0.64\textwidth}
        \centering
        \includegraphics[height=0.5\textwidth]{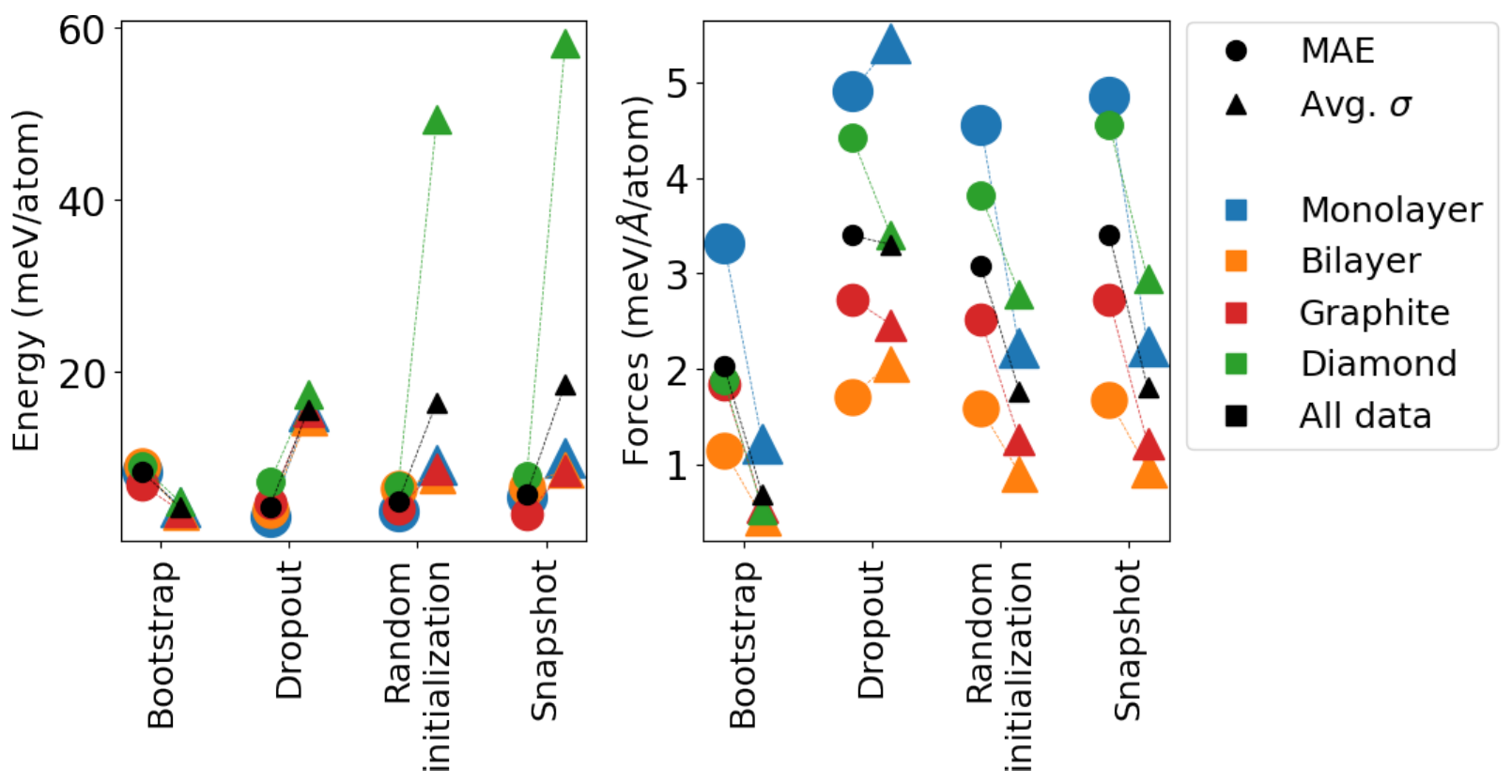}
        \caption{}
	\label{fig:error_uncertainty}
    \end{subfigure}
    \caption[Pearson correlation coefficients, the mean absolute error, and average uncertainty of the energy and forces in the test set]{
	(a) Pearson correlation coefficients between the residual and the uncertainty in the test set.
	The columns represent different ensemble models, while the rows represent different structure types in the dataset, with the last row gives the correlation coefficient calculated using the entire data points.
	(b) The mean absolute error (MAE) and average uncertainty of the energy and forces in the test set.
    }
    \label{fig:accuracy_energy_forces_correlation}
\end{figure*}

From this result, the distinction between sample distributions is more pronounced across structure types than across ensemble models.
To further compare the ensemble methods, we compute the Pearson correlation coefficients between residuals and uncertainties (Fig.~\ref{fig:correlation}), as well as the mean absolute error (MAE) and average uncertainty of the energy and force predictions (Fig.~\ref{fig:error_uncertainty}), over the test set for each ensemble model and structure type.
Since the energy and force contributions are analyzed separately, the data points no longer need to be weighted differently, which justifies the use of the MAE instead of the average residual.
We first note that the random initialization and snapshot ensembles exhibit very similar performance.
This is evident from both the correlation values and the MAE versus average uncertainty plots for energy and forces across all considered carbon allotropes.
However, the snapshot ensemble is significantly more computationally efficient because it requires training only a single model.

The correlation analysis further shows that the residuals and uncertainties are most strongly correlated for the bootstrap ensemble, followed by the dropout ensemble.
Despite this strong correlation, however, the bootstrap ensemble often produces overconfident predictions, particularly for the force quantities.
In contrast, the dropout ensemble yields average uncertainties that either exceed or closely match the MAEs, suggesting better-calibrated uncertainty estimates.
Furthermore, the resulting uncertainties provide more conservative bounds when interpreted as estimates of the actual prediction error.
These observations suggest a potential preference for the dropout ensemble over the bootstrap ensemble for uncertainty quantification.

Additionally, the overconfident behavior of the bootstrap ensemble likely arises because the training data consists of non-independent snapshots from MD trajectories.
With such data, certain regions of the distribution are overrepresented.
As a result, bootstrap resampling does not introduce sufficient variability into the training sets, leading to an underestimated spread in ensemble predictions and, consequently, overconfident uncertainty estimates.
In contrast, the dropout ensemble introduces variability through stochastic masking during training and inference, making it less sensitive to the non-independence of the training data.

That said, the overall differences in performance between the ensemble methods remain minimal within this ID domain.
Except for the bootstrap case, the correlations between residuals and uncertainties are generally weak, and the MAE and average uncertainty are of similar magnitudes across methods.
Therefore, this analysis alone does not provide sufficient evidence to definitively identify the best-performing ensemble model.

\subsection{Out-of-distribution comparison}
\label{sec:out-of-distribution}

We further extend the ensemble models comparison analysis in the OOD domain, given by some large-scale material properties.
As proxies for large-scale properties, we compare the energy cold curves and phonon dispersion relations for graphene, graphite, and diamond structures.
These OOD results expose nontrivial and, in some cases, counterintuitive uncertainty behavior across all ensemble methods that is not evident from in-distribution metrics alone; a detailed discussion is provided in Sec.~\ref{sec:discussion}.

\begin{figure*}[!hbt]
    \centering
    \includegraphics[width=0.9\textwidth]{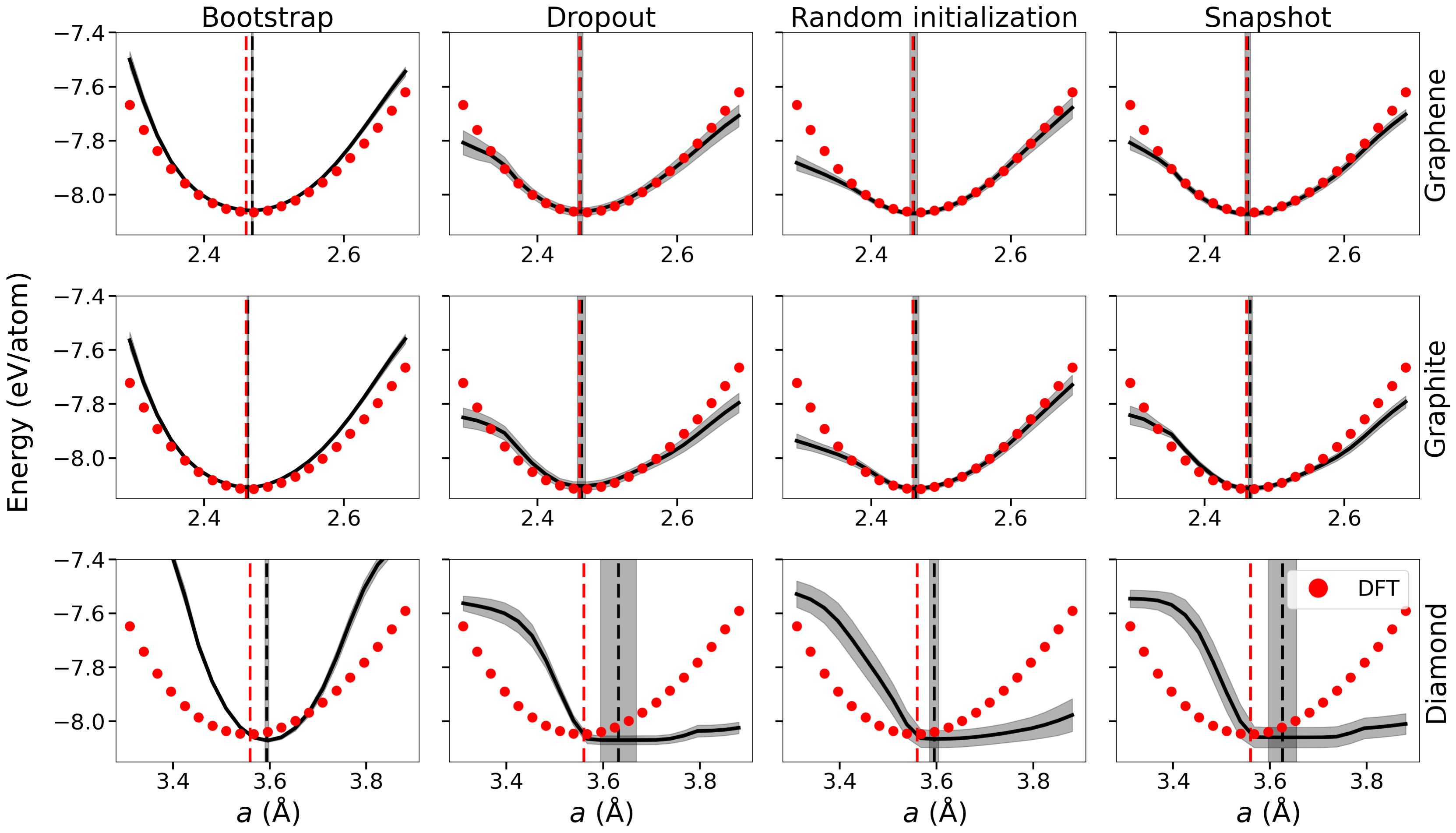}
    \caption[Energy cold curve predictions using different ensemble models]{
	Energy cold curve predictions for (top) graphene, (middle) graphite, and (bottom) diamond structures using different ensemble models.
	The ensemble-averaged predictions are shown as black curves, with the grey envelope representing the one-standard-deviation uncertainty.
	DFT ground truth values (red dots) are overlaid for comparison with the predicted values.
	For each structure, the lattice parameter is perturbed by $\pm 10\%$ relative to the reference equilibrium lattice constant from the Materials Project repository (red vertical dashed line in each panel).
	The predicted equilibrium lattice constant (black vertical dashed line) with its associated one-standard-deviation uncertainty is also shown in each panel.
    }
    \label{fig:energy_cold_curve}
\end{figure*}

Figure~\ref{fig:energy_cold_curve} shows the energy cold curve predictions for the three carbon allotropes considered.
The ensemble mean predictions are plotted as black curves, with one-standard-deviation uncertainty indicated by the gray envelope.
The corresponding equilibrium lattice constant predictions are indicated by vertical dashed lines, with the gray envelope also reflecting the uncertainty in these predicted equilibrium positions.
DFT reference values are overlaid to assess the accuracy of the predictions.
For graphene and graphite, the predicted energies near equilibrium are sufficiently accurate and precise, with the DFT values falling within the uncertainty bounds.
However, the predictions begin to deviate under large compression or tension, and in many cases, the uncertainty does not grow rapidly enough to reflect the increasing prediction error.

In contrast, the predictions for the diamond structure are significantly less accurate, and most of the DFT reference values fall outside the predicted uncertainty bounds, indicating overconfident predictions.
Furthermore, the models also fail to capture the expected parabolic shape of the energy curve near equilibrium.
In particular, in the extension regime, the predicted energy flattens instead of increasing, diverging substantially from the expected physical behavior.
Combined with the fact that the uncertainty estimates remain low, this result suggests that the ensemble models are confidently making incorrect predictions, which can lead to misleading conclusions in downstream material property predictions.
A possible explanation for this failure is discussed in Sec.~\ref{sec:discussion} in terms of model extrapolation beyond the training data.

Phonon dispersion predictions follow similar trends to the energy cold curves: reasonably accurate for graphene and graphite but much less so for diamond.
For graphene and graphite, most phonon branches are reproduced within the uncertainty bounds, with minor underestimation in certain modes, such as the flexural optical mode near the $\Gamma$ point.
In contrast, the diamond spectra show large deviations from DFT, including underestimated optical branches and noisy acoustic modes.
The wide uncertainty bounds fail to capture the true values, indicating unreliable uncertainty estimates.
Further details and figures are provided in the supplementary material.


\section{Discussion}
\label{sec:discussion}

\subsection{Feature space analysis}
\label{sec:pca}

\begin{figure}[!hbt]
    \centering
    \includegraphics[width=0.45\textwidth]{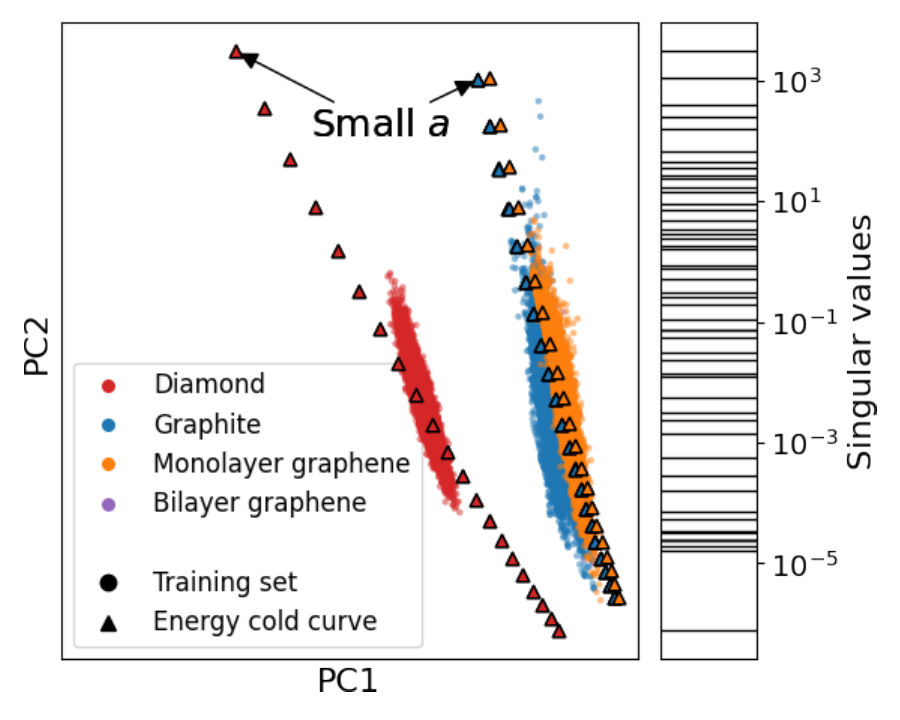}
    \caption[PCA embedding of local atomic environments in the training data and in the energy cold curve predictions]{
	PCA embedding of local atomic environments in the training data (clusters of points) and in the energy cold curve predictions (triangles), along with the singular values obtained from the SVD of the stacked descriptor matrix used in the PCA analysis.
	Different colors indicate different crystal structures (the bilayer graphene cluster is beneath the graphite and monolayer clusters).
	The embedding space shown is defined by the two most dominant principal components corresponding to the two largest singular values.
	These dominant components capture 98\% of the variance in the training data, calculated as the fraction of variance explained by the sum of their squared singular values.
	This PCA embedding plot illustrates that the energy cold curve calculations involve extrapolation beyond the training data, particularly in the case of diamond.
    }
    \label{fig:pca}
\end{figure}

Neural network models are black-box predictors that lack explicit physical constraints, and they are known to exhibit degraded accuracy in extrapolation regimes.
The decline in prediction accuracy observed in our ensemble models---particularly for diamond---can be attributed to this limitation.
To investigate this, we perform a principal component analysis (PCA) of the local atomic environments to assess the coverage of the training data and visualize the relationship between training and evaluation points.
We fit a PCA model to the atomic environment representations from the training set, then project these representations onto the subspace defined by the two most dominant principal components.
Together, these two components capture 98\% of the dataset’s variance, computed as the ratio of the sum of their squared singular values to the total sum of squared singular values.
The resulting embeddings are shown as clusters in Fig.~\ref{fig:pca}, where each point corresponds to an atomic environment and is colored by crystal structure.
For comparison, we also overlay the embeddings of atomic environments sampled during the energy cold curve calculations.

From the PCA plot, we observe that atomic environments corresponding to highly compressed graphene and graphite structures lie outside the main cluster of training data, indicating extrapolation.
Notably, these configurations correspond to regions where the energy cold curve predictions deviate from the DFT ground truth across all ensemble models, suggesting that the reduced accuracy may be linked to the model's operation in extrapolative regimes.
This relationship is further supported by the observation for diamond, where there is minimal overlap between the training data and the atomic environments involved in the cold curve calculations.
The lack of representation in the training set contributes to substantially larger prediction errors in the diamond energy values.

\subsection{Dropout Ratio and Uncertainty}
\label{sec:dropout-ratio}

While extrapolation appears to be linked to reduced prediction accuracy, it does not fully explain the overconfidence observed in the model predictions.
Although small-scale predictions remain well-calibrated (see Fig.~\ref{fig:accuracy_energy_forces_correlation}), the downstream energy predictions in the extrapolation regime are notably overconfident (see Fig.~\ref{fig:energy_cold_curve}).
In these cases, the estimated uncertainties underestimate the actual prediction errors, making uncertainty an unreliable approximation of prediction error.

\begin{figure*}[!hbt]
    \centering
    \includegraphics[width=0.9\textwidth]{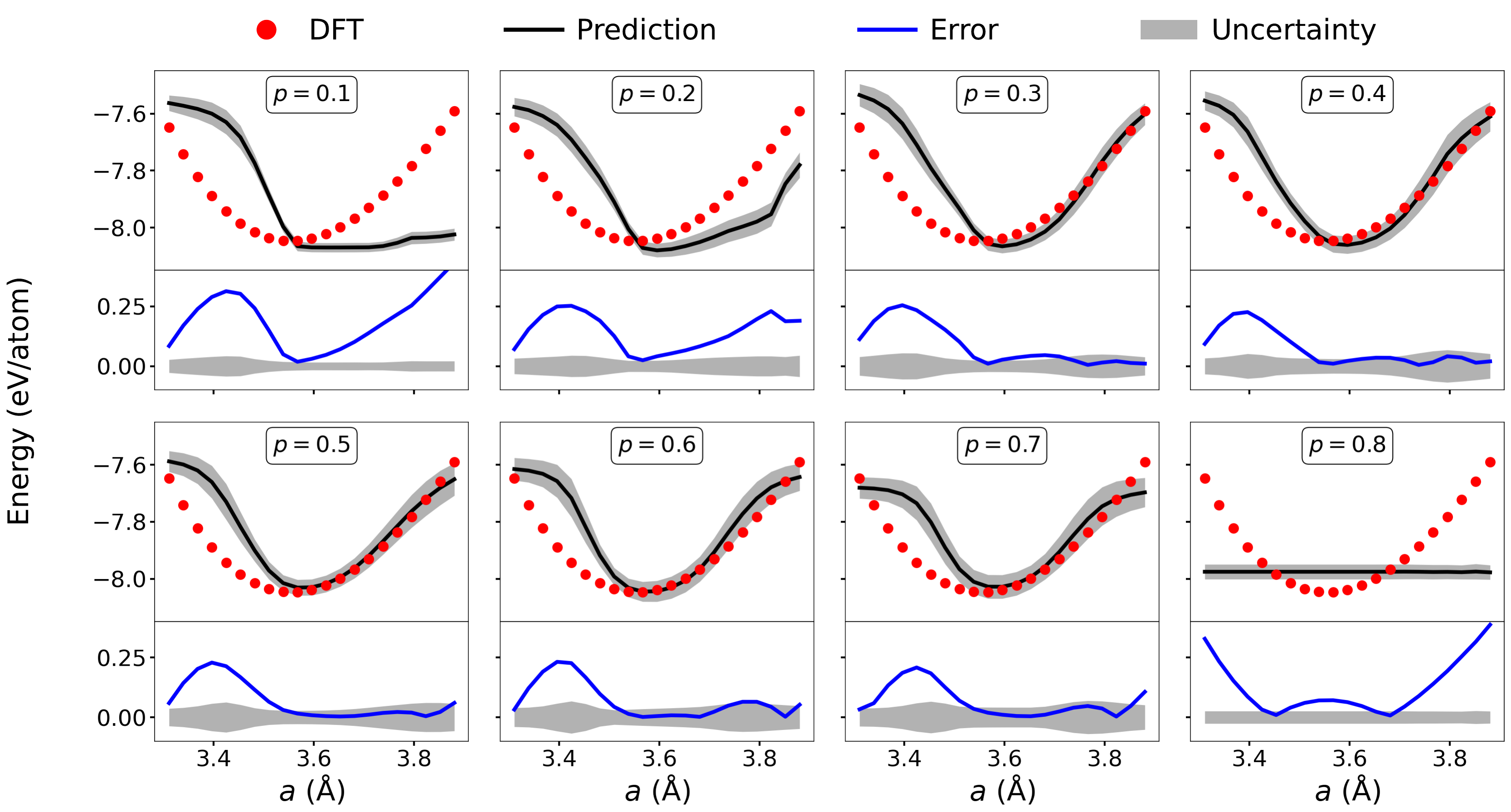}
    \caption[Energy cold curve predictions and uncertainties for diamond with several dropout ratios]{
	Energy cold curve predictions and associated uncertainties for diamond at varying dropout ratios.
	The lower part of each panel shows the prediction error relative to the DFT ground truth alongside the predicted uncertainty.
	This visualization highlights how uncertainty increases with higher dropout ratios, particularly in the tension regime, and how the prediction error correlates with uncertainty.
	Interestingly, in the compression regime, both the error and uncertainty remain largely unchanged.
	Additionally, at very high dropout ratios, the model produces nearly constant predictions with low variability, suggesting a loss of learning capacity.
    }
    \label{fig:energy_cold_curve_dropout_diamond}
\end{figure*}

One strategy to mitigate overconfident predictions is to tune hyperparameters within each ensemble UQ method to increase variability among ensemble members.
For instance, increasing the dropout ratio in dropout ensembles introduces greater diversity across models, which in turn yields larger prediction uncertainty.
This effect is illustrated in Fig.~\ref{fig:energy_cold_curve_dropout_diamond}, where we compare the energy cold curve predictions and associated uncertainties for the diamond structure at different dropout ratios (results for hexagonal structures are included in the supplementary material).
As the dropout ratio increases from the baseline value of $p=0.1$, the uncertainty bands widen and the model becomes more cautious in its predictions.

However, this comes at the cost of reduced learning capacity. When the dropout ratio becomes too high---for example, 80\%---the network is unable to learn relevant features, and the predicted energy curves become unphysical (e.g., appearing flat where positive curvature is expected).
Unfortunately, there is no universal guideline for selecting an optimal dropout ratio.
While some prior studies suggest values around 50\% \cite{baldi_understanding_2013,srivastava_dropout_2014}, the choice remains largely empirical and is likely sensitive to both network architecture and task complexity.
This also points to an important direction for future work: whether we should design wider and/or deeper networks that can better tolerate large dropout ratios.

Additionally, while increasing the dropout ratio appears to improve energy prediction accuracy in the tension regime for our case study, this outcome should not be interpreted as generalizable.
In our example, the model begins to exhibit the correct physical behavior---rising energy under lattice tension---as the dropout ratio increases.
However, this improvement seems incidental rather than systematic.
As a counterexample, in the compression regime, increasing the dropout ratio has little effect on prediction accuracy; the energy errors remain largely unchanged.

These contrasting outcomes suggest that tuning the dropout ratio should not be viewed as a reliable method for improving model accuracy.
The observed improvements are highly context-dependent and cannot be expected to generalize across different systems or regimes.
However, developing approaches to improve predictive accuracy is beyond the scope of this study.
Instead, our focus is on improving uncertainty calibration---specifically, using dropout tuning to make predictive uncertainties more aligned with actual model errors.
In this regard, increasing the dropout ratio proves effective for mitigating overconfidence by inflating uncertainty to better align it with model errors.

\subsection{Uncertainty Beyond Training Data}
\label{sec:uncertainty-extrapolation}

\begin{figure*}[!hbt]
    \centering
    \includegraphics[width=0.9\textwidth]{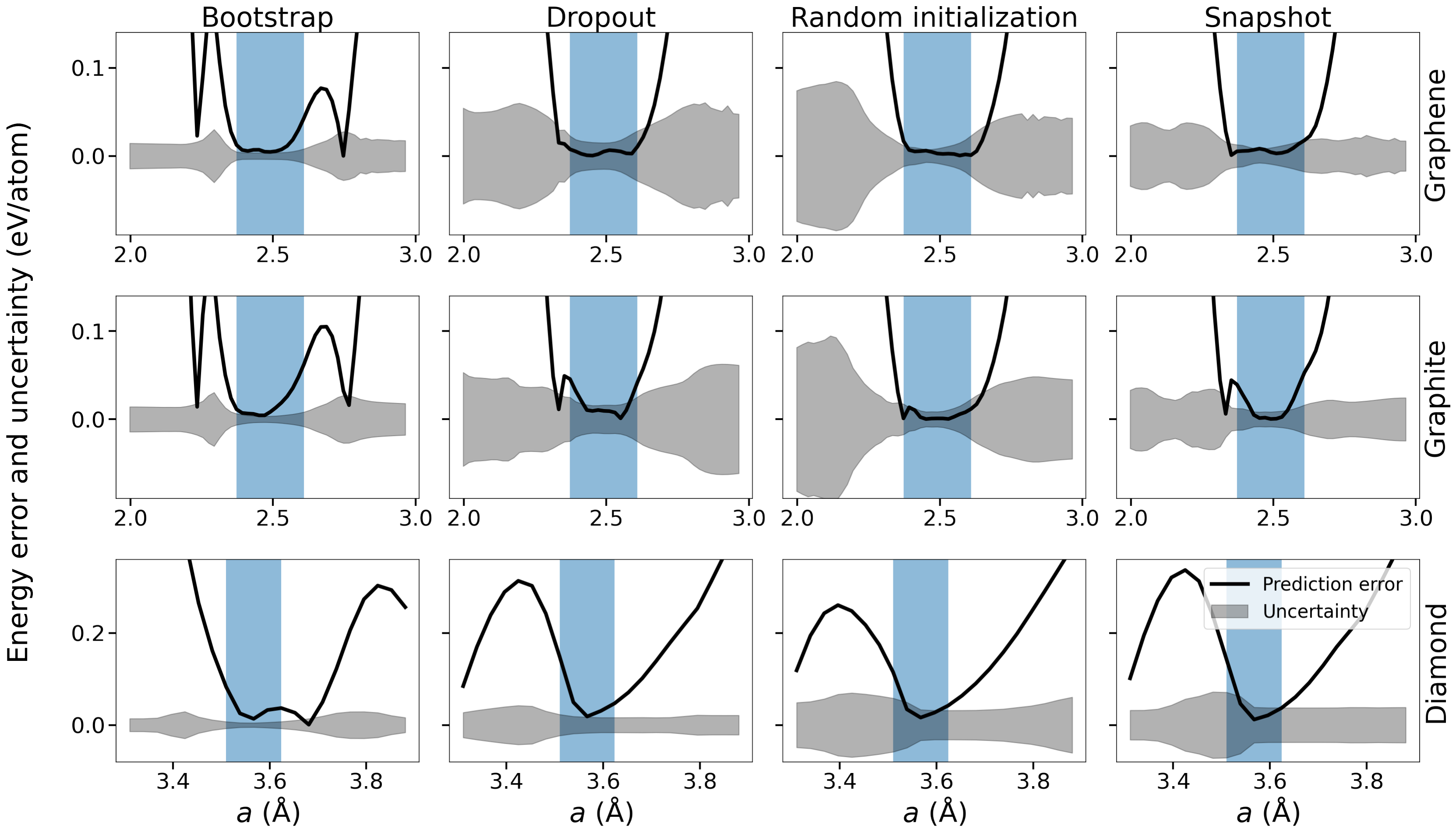}
    \caption[Energy cold curve prediction errors and uncertainties]{
	Prediction errors (black curves) and associated uncertainty estimates (gray envelopes) for energy cold curve predictions.
	The blue-shaded regions indicate the interpolation domain, as determined from the feature space analysis in Fig.~\ref{fig:pca}.
	This figure highlights a counterintuitive uncertainty behavior, where the predictions uncertainties decrease as the ensemble models extrapolate beyond the training data.
    }
    \label{fig:energy_cold_curve_extrapolation}
\end{figure*}

In addition to producing overconfident predictions in the compression regime, the uncertainty estimates in Fig.~\ref{fig:energy_cold_curve_dropout_diamond} reveal a counterintuitive trend.
The prediction uncertainty, rather than increasing with extrapolation, instead decreases under extreme compression.
A similar pattern appears in the tension regime for certain dropout ratios, where the model is again pushed far beyond the training distribution.
This behavior is not limited to dropout ensembles; it consistently appears across all ensemble models and carbon allotropes, as shown in Fig.~\ref{fig:energy_cold_curve_extrapolation}, with the blue region indicating the interpolation domain.
While uncertainty initially rises modestly as the model begins to extrapolate---consistent with previous findings \cite{wen_uncertainty_2020,carrete_deep_2023}---it does so at a lower rate than the prediction error and eventually plateaus or even declines, despite increasing inaccuracy.
For the hexagonal structures, we do need to extend the lattice parameter perturbation to $\pm 20\%$ in order to observe this phenomenon.

\begin{figure*}[!hbt]
    \centering
    \includegraphics[width=0.9\textwidth]{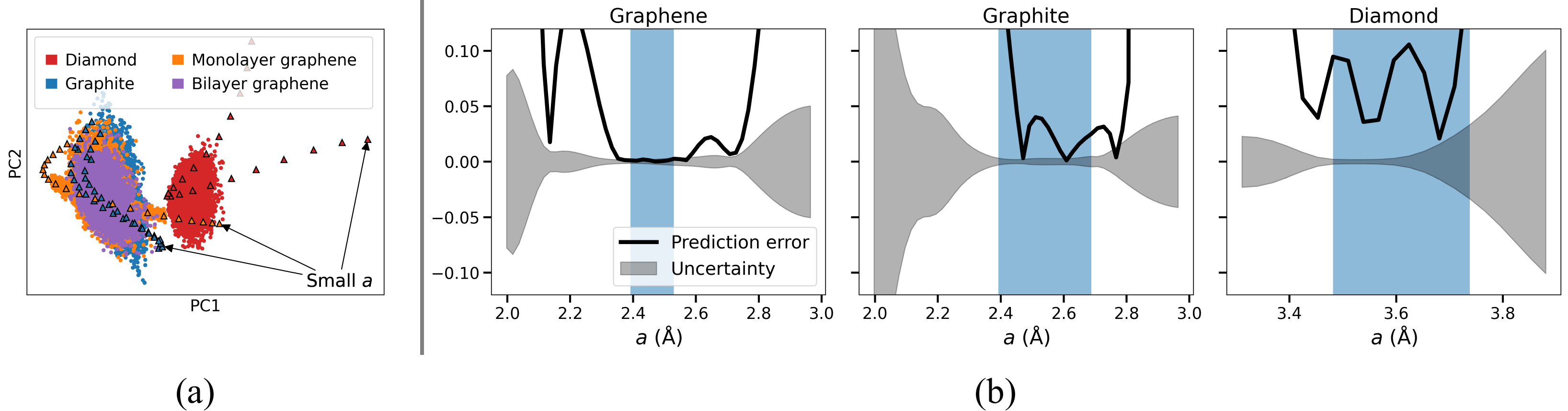}
    \caption[Schnet training data PCA and energy cold curve predictions]{
	(a) PCA embedding of the graph representation of local atomic environments used by SchNet, comparing the training data coverage with the atomic environments sampled in the OOD energy cold curve predictions.
        (b) Prediction errors and associated uncertainty estimates for energy cold curve predictions of graphene, graphite, and diamond structures.
        These results show that overconfident and counterintuitive uncertainty behavior---such as decreasing or plateauing uncertainty under extreme extrapolation---is also observed in a modern graph neural network potential.
        This indicates that these issues are not specific to a particular model architecture and may occur across different NN models
    }
    \label{fig:energy_cold_curve_schnet}
\end{figure*}

This counterintuitive uncertainty behavior is not limited to the NNIP architecture used in this study.
Similar trends are observed with graph neural network (GNN)--based models, as illustrated by SchNet results in Fig.~\ref{fig:energy_cold_curve_schnet}.
The SchNet potential was trained on the same dataset for 4,000 epochs to achieve comparable ID errors to the NNIP ensembles.
Even with this modern GNN model, prediction uncertainties remain overconfident when the model extrapolates, and in some cases the uncertainties plateau as extrapolation increases.
While these observations do not imply a general rule for uncertainty behavior under extrapolation, they indicate that this phenomenon is not specific to a single architecture and may occur across different NN models.

These findings further reveal a key limitation of ensemble-based UQ methods.
While uncertainty can serve as a signal that extrapolation is occurring, it does not reliably indicate how far the model has extrapolated beyond the training domain.
In our controlled experiments, we can trace this progression by systematically perturbing the lattice parameter and evaluating the prediction errors.
However, in practical applications, such ground truth is rarely available, especially for large-scale or high-dimensional simulations.
Consequently, models may exhibit low uncertainty even while operating far beyond their training domain.
This situation is inherently ambiguous: on one hand, the model may still be interpolating and making accurate predictions; on the other hand, it may be confidently producing inaccurate results in regions far from the training data.
Post-hoc calibration may improve the numerical interpretation of the uncertainty estimates, but it does not necessarily resolve this ambiguity under distribution shift \cite{dale_when_2025}.
Adding to the challenge, the boundary between interpolation and extrapolation is often subtle and difficult to delineate in high-dimensional feature spaces, limiting the effectiveness of uncertainty as a standalone indicator of model reliability.

\begin{figure}[!hbt]
    \centering
    \includegraphics[width=0.45\textwidth]{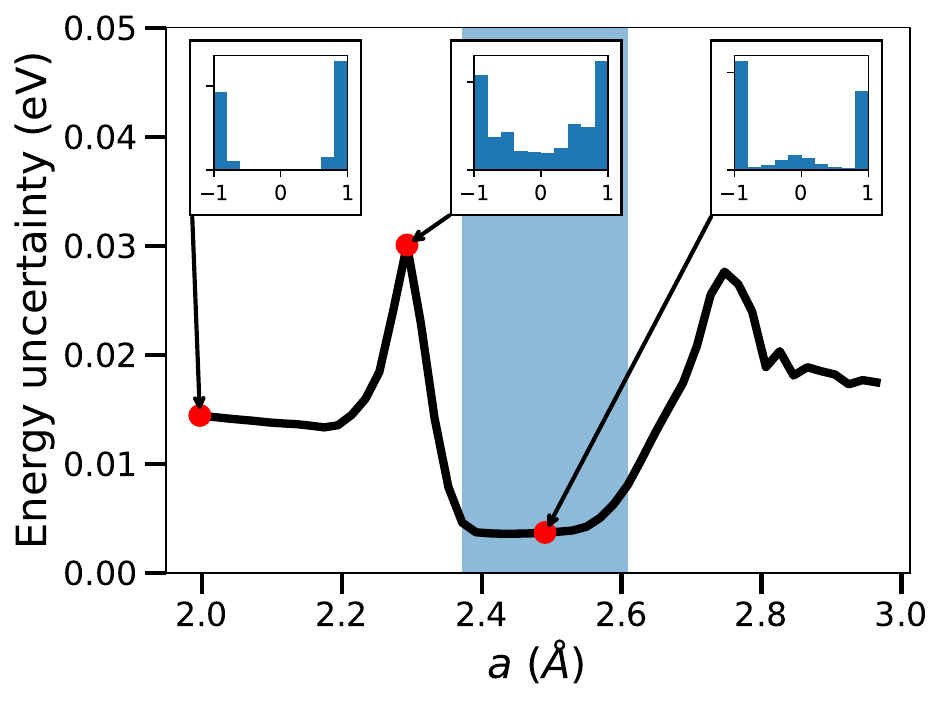}
    \caption[Energy uncertainty with the distributions of some outputs of the final activation function]{
	Energy uncertainty estimated from the bootstrap ensemble for graphene, along with the distributions of the final activation function outputs from the NN model evaluated at selected lattice parameters.
    }
    \label{fig:saturation}
\end{figure}

There are several possible explanations for the counterintuitive drop in uncertainty observed during extreme extrapolation.
One hypothesis is that using the hyperbolic tangent activation function---a bounded function---causes model outputs to saturate in regions far from the training data.
This saturation could reduce the model’s sensitivity to input variations, leading to lower estimated uncertainty.
To test this hypothesis, we examined the output of the final activation function for atomic configurations sampled from the energy cold curve calculation.
In Fig.\ref{fig:saturation}, we show the predicted energy uncertainty for graphene’s cold curve via a bootstrap ensemble (cf. the gray band in the top-left plot of Fig.\ref{fig:energy_cold_curve_extrapolation}) along with the distributions of final activation outputs at three lattice parameters.
For the far extrapolation case ($a = 2.0~\text{\AA}$), the outputs are clearly saturated at the hyperbolic tangent’s extremes, which may account for the reduced uncertainty.
At $a = 2.3~\text{\AA}$, where uncertainty is relatively high, the outputs are more dispersed, consistent with the hypothesis.
However, near the equilibrium configuration, the hypothesis breaks down: despite being within the training regime, the activation outputs also appear saturated at the same extreme values.
These observations suggest that activation function saturation alone does not fully explain the drop in uncertainty during extrapolation.
While indicative, these findings are not conclusive, and further investigation---such as testing alternative activation functions---is necessary for a more definitive understanding.

Another possible explanation is that in these extreme extrapolation regions, the ensemble of models lacks sufficient information to produce meaningful variability in their predictions, causing the outputs to collapse and underestimate uncertainty.
Furthermore, this phenomenon may reflect a fundamental mismatch in the types of uncertainty captured.
Ensemble methods primarily quantify epistemic uncertainty, but in extreme extrapolation, the model’s epistemic uncertainty may be underestimated or poorly represented.
Incorporating distance-based measures may help improve how epistemic uncertainty is quantified in these regimes.
Additionally, aleatoric uncertainty is typically not captured by these methods, though it usually plays a secondary role in extrapolation errors.
While a deeper investigation into these aspects lies beyond the scope of this study, we refer the interested reader to related works that analyze and compare epistemic and aleatoric uncertainties, as they may offer further insight into this issue \cite{busk_calibrated_2021,heid_characterizing_2023,wimer_benchmarking_2026}.


\section{Conclusion}
\label{sec:conclusion}

In this study, we investigate when predictive uncertainty can serve as a reliable proxy for model accuracy.
We conduct a comparative analysis of several ensemble-based UQ methods for NNIPs, including bootstrap, dropout, random initialization, and snapshot ensembles.
This comparison allows us to identify key factors that influence the relationship between prediction precision and its accuracy.
The predictive performance and associated uncertainties are evaluated in both ID and OOD settings, represented by the energy cold curve and phonon dispersion relations predictions for various carbon allotropes.

Our results indicate that prediction precision serves as a reliable proxy for accuracy within the ID domain.
In the OOD domain, however, this relationship holds only when the model remains in an interpolative regime; once it begins to extrapolate beyond the training domain, predictions often become overconfident.
Although fine-tuning ensemble hyperparameters---such as the dropout rate---can reduce overconfidence to some extent, this strategy has limited effectiveness and should not be relied upon to improve accuracy.
More surprisingly, we observe a counterintuitive trend in uncertainty behavior during extrapolation, where the estimated uncertainty often plateaus or even declines instead of continuing to rise with increasing prediction error.
We consider two possible explanations for this behavior, including activation function saturation and a mismatch between the types of uncertainty being captured (epistemic vs. aleatoric).
While we conduct some preliminary tests related to these hypotheses, the results reveal counterexamples that cast doubt on them, suggesting they do not fully explain the anomaly.
Nevertheless, further investigation is needed to reach more definitive conclusions.
Furthermore, while the present analysis focuses on NNIP architecture, extending this study to a broader range of MLIP architectures remains an important direction for future work.

In the meantime, a more practical strategy is to minimize model extrapolation whenever possible.
One approach is to analyze the feature space of the training data to identify regions that are poorly represented \cite{karabin_entropy-maximization_2020,montes_de_oca_zapiain_training_2022,schwalbe-koda_model-free_2024}.
Another complementary strategy is to apply active learning techniques that enable the model to query or prioritize the most informative data points \cite{behler_representing_2014,gubaev_accelerating_2019,kurniawan_information-matching_2026}.
While these are not novel suggestions in the context of NNIPs, we reiterate their importance here in the specific context of uncertainty estimation.

Unfortunately, in the absence of a coherent theory of learning for NN models, it remains difficult to fully explain or resolve the behavior of uncertainty estimates, especially in extrapolation.
Without deeper theoretical insight, our ability to diagnose or correct these issues is inherently limited.
While UQ studies for NN models can still provide valuable perspectives, we must approach their conclusions with caution.
Until we develop a better understanding of why NN models work and how they generalize, their uncertainty estimates should be treated with skepticism, especially in high-stakes or extrapolative scenarios.


\ack{
    This work was supported by the National Science Foundation under Awards Nos. DMR-1834251 and DMR-1834332.
    The calculations were done on computational facilities provided by the Brigham Young University Office of Research Computing.
    We also thank Nicholas Wimer, Fei Zhou, Amit Gupta, Ilia Nikiforov, Juliane M\"{u}ller, and Vincenzo Lordi for valuable discussions and insights.
}

\data{
    The code and data supporting the findings of this study are available at the following GitHub repository:
    \href{https://github.com/yonatank93/compare_UQ}{https://github.com/yonatank93/compare\_UQ}.
}

\roles{
    Yonatan Kurniawan: Conceptualization, formal analysis, investigation, methodology, visualization, and writing---original draft.
    Mingjian Wen: Data curation, methodology, and writing---review and editing.
    Ellad B. Tadmor: Project administration, funding acquisition, writing---review and editing.
    Mark K. Transtrum: Conceptualization, formal analysis, methodology, supervision, and writing---review and editing.
}

\suppdata{
    The supplementary material includes an introduction to atomic descriptors, with emphasis on the atom-centered symmetry functions (ACSFs) used in this study.
    It also provides additional UQ results for large-scale properties, including energy cold curves for graphene and graphite structures evaluated at different dropout ratios, and phonon dispersion relations for three carbon allotropes across different ensemble methods and dropout ratios.
}

\bibliographystyle{iopart-num}
\bibliography{refs,refs_zotero}

\end{document}


\maketitle 

\tableofcontents

\section{Atom-centered symmetry function}
\label{sec:descriptor}

Raw atomic coordinates in an atomic configuration cannot be directly used as input to neural network interatomic potentials (NNIPs) for several fundamental reasons.
First, the number of neighboring atoms can vary from one atom to another, resulting in coordinate vectors of inconsistent lengths, which standard neural networks are not equipped to handle.
Second, and more importantly, raw coordinates do not preserve essential physical symmetries: they are not invariant under rigid-body translations, rotations, or permutations of atoms of the same type.
To address these challenges, the structural and chemical information of atoms and their neighborhoods is encoded into fixed-length feature vectors known as atomic descriptors.
These descriptors aggregate information from an atom's local environment, regardless of the number of neighbors, while explicitly enforcing the symmetries required for physically meaningful energy predictions.
Moreover, the choice of atomic descriptor can significantly influence the accuracy and generalizability of NNIPs, as it determines which features of the atomic environment are accessible to the learning algorithm.

In this work, we use atom-centered symmetry functions (ACSFs) \cite{behler_generalized_2007_edited,behler_atom-centered_2011_edited,artrith_high-dimensional_2012_edited} to construct the descriptor vectors.
The radial and angular components of the environment surrounding atom $n$ are defined as
\begin{align}
  \label{eq:acsf_desc_2}
  G_i^2 &= \sum_{j \neq i}^N e^{-\eta_2 (r_{ij} - R_s)^2} f_c(r_{ij}), \\
  \label{eq:acsf_desc_3}
  G_i^3 &= 2^{1-\zeta} \sum_{j \neq i}^N \sum_{\substack{k > j \\ k \neq i}}^N (1 + \lambda \cos \theta_{jik})^{\zeta}
  e^{-\eta_3 (r_{ij}^2 + r_{ik}^2 + r_{jk}^2)} f_c(r_{ij}) f_c(r_{ik}) f_c(r_{jk}),
\end{align}
where $r_{ij}$ denotes the distance between atoms $i$ and $j$, and $\theta_{jik}$ is the angle formed by the bonds $i-j$ and $i-k$.
A smooth cutoff function $f_c$, defined as
\begin{equation}
    \label{eq:cutoff}
    f_c(r) =
    \begin{cases}
      \frac{1}{2} \left( \cos\left( \frac{\pi r}{R_c} \right) + 1 \right), & r \leq R_c \\
      0, & r > R_c
    \end{cases},
\end{equation}
ensures locality by gradually reducing the descriptor contribution to zero beyond the cutoff radius $R_c$, which is set to 5~\AA~in this study.

The descriptor vector for each atomic environment is constructed by evaluating Eqs.~\eqref{eq:acsf_desc_2} and \eqref{eq:acsf_desc_3} using multiple sets of hyperparameter values, listed in Tables~\ref{tab:hyperparam_g2} and \ref{tab:hyperparam_g3}.
To ensure consistent scaling across features, each descriptor component is then normalized by subtracting its mean and dividing by its standard deviation, both computed over the training set.
The resulting standardized descriptors are then used as input to the NNIP.

\begin{table}[!h]
    \centering
    \begin{tabular}{c @{\hspace{2em}} c @{\hspace{2em}} c @{\hspace{2em}}}
      \hhline{===}
      \rule{0pt}{2.5ex} No. & $\eta_2$ (Bohr$^{-2}$) & $R_s$ (Bohr) \\
      \hline
      1 & $0.001   $ & $0$ \\
      2 & $0.01\phz$ & $0$ \\
      3 & $0.02\phz$ & $0$ \\
      4 & $0.035   $ & $0$ \\
      5 & $0.06\phz$ & $0$ \\
      6 & $0.1\phzz$ & $0$ \\
      7 & $0.2\phzz$ & $0$ \\
      8 & $0.4\phzz$ & $0$ \\
      \hhline{===}
    \end{tabular}
    \caption{Hyperparameters for the radial part of the atom-centered symmetry functions descriptor}
    \label{tab:hyperparam_g2}
\end{table}

\begin{table*}[!h]
    \centering
    \resizebox{\textwidth}{!}{
      \begin{tabular}{
        c @{\hspace{1.5em}} c @{\hspace{1.5em}} c @{\hspace{1.5em}} c @{\hspace{1.5em}} |
        c @{\hspace{1.5em}} c @{\hspace{1.5em}} c @{\hspace{1.5em}} c @{\hspace{1.5em}} |
        c @{\hspace{1.5em}} c @{\hspace{1.5em}} c @{\hspace{1.5em}} c @{\hspace{1.5em}}
        }
        \hhline{====|====|====}
        \rule{0pt}{2.5ex} No. & $\zeta$ & $\lambda$ & $\eta_3$ (Bohr$^{-2}$) & No. & $\zeta$ & $\lambda$ & $\eta_3$ (Bohr$^{-2}$) & No. & $\zeta$ & $\lambda$ & $\eta_3$ (Bohr$^{-2}$)\\
        \hline
        ~1 & $\phz1$  & $   -1$ & $0.0001   $ & 17 & $\phz4$ & $   -1$ & $0.015$ & 33 & $\phz4$ & $   -1$ & $0.045   $ \\
        ~2 & $\phz1$  & $\phm1$ & $0.0001   $ & 18 & $\phz4$ & $\phm1$ & $0.015$ & 34 & $\phz4$ & $\phm1$ & $0.045   $ \\
        ~3 & $\phz2$  & $   -1$ & $0.0001   $ & 19 & $   16$ & $   -1$ & $0.015$ & 35 & $   16$ & $   -1$ & $0.045   $ \\
        ~4 & $\phz2$  & $\phm1$ & $0.0001   $ & 20 & $   16$ & $\phm1$ & $0.015$ & 36 & $   16$ & $\phm1$ & $0.045   $ \\
        ~5 & $\phz1$  & $   -1$ & $0.003\phz$ & 21 & $\phz1$ & $   -1$ & $0.025$ & 37 & $\phz1$ & $   -1$ & $0.08\phz$ \\
        ~6 & $\phz1$  & $\phm1$ & $0.003\phz$ & 22 & $\phz1$ & $\phm1$ & $0.025$ & 38 & $\phz1$ & $\phm1$ & $0.08\phz$ \\
        ~7 & $\phz2$  & $   -1$ & $0.003\phz$ & 23 & $\phz2$ & $   -1$ & $0.025$ & 39 & $\phz2$ & $   -1$ & $0.08\phz$ \\
        ~8 & $\phz2$  & $\phm1$ & $0.003\phz$ & 24 & $\phz2$ & $\phm1$ & $0.025$ & 40 & $\phz2$ & $\phm1$ & $0.08\phz$ \\
        ~9 & $\phz1$  & $   -1$ & $0.008\phz$ & 25 & $\phz4$ & $   -1$ & $0.025$ & 41 & $\phz4$ & $   -1$ & $0.08\phz$ \\
        10 & $\phz1$  & $\phm1$ & $0.008\phz$ & 26 & $\phz4$ & $\phm1$ & $0.025$ & 42 & $\phz4$ & $\phm1$ & $0.08\phz$ \\
        11 & $\phz2$  & $   -1$ & $0.008\phz$ & 27 & $   16$ & $   -1$ & $0.025$ & 43 & $   16$ & $\phm1$ & $0.08\phz$ \\
        12 & $\phz2$  & $\phm1$ & $0.008\phz$ & 28 & $   16$ & $\phm1$ & $0.025$ &&&& \\
        13 & $\phz1$  & $   -1$ & $0.015\phz$ & 29 & $\phz1$ & $   -1$ & $0.045$ &&&& \\
        14 & $\phz1$  & $\phm1$ & $0.015\phz$ & 30 & $\phz1$ & $\phm1$ & $0.045$ &&&& \\
        15 & $\phz2$  & $   -1$ & $0.015\phz$ & 31 & $\phz2$ & $   -1$ & $0.045$ &&&& \\
        16 & $\phz2$  & $\phm1$ & $0.015\phz$ & 32 & $\phz2$ & $\phm1$ & $0.045$ &&&& \\
        \hhline{====|====|====}
      \end{tabular}
    }
    \caption{Hyperparameters for the angular part of the atom-centered symmetry functions descriptor}
    \label{tab:hyperparam_g3}
\end{table*}

\newpage
\section{Other results}
\label{sec:other-results}

This section presents additional prediction and uncertainty results for the energy cold curve and phonon dispersion relations that were not included in the main text.
These results were omitted for brevity but provide further support for the analyses presented in the main manuscript.
Section~\ref{sec:energy-cold-curve} provides additional comparisons of energy cold curve predictions and their associated uncertainties across different dropout ratios for graphene and graphite.
Section~\ref{sec:phonon-dispersion} presents phonon dispersion predictions and uncertainties obtained using different UQ methods. This section also includes comparisons of phonon dispersion results across varying dropout ratios.

\subsection{Energy cold curve }
\label{sec:energy-cold-curve}

Here, we present additional comparisons of energy cold curve predictions and their associated uncertainties for graphene and graphite structures, evaluated across different dropout ratios.
The list of lattice parameters is generated by perturbing the equilibrium lattice constant by $\pm~10\%$.
As expected, the prediction uncertainties increase with the dropout ratio, reducing overconfidence especially near the edges of the energy curve, where the model begins to extrapolate beyond the training regime.
However, excessively large dropout ratios (e.g., $p \geq 0.8$) can lead to a catastrophic failure in model performance, likely due to reduced learning capacity.

\begin{figure*}[!hbt]
    \centering
    \includegraphics[width=0.9\textwidth]{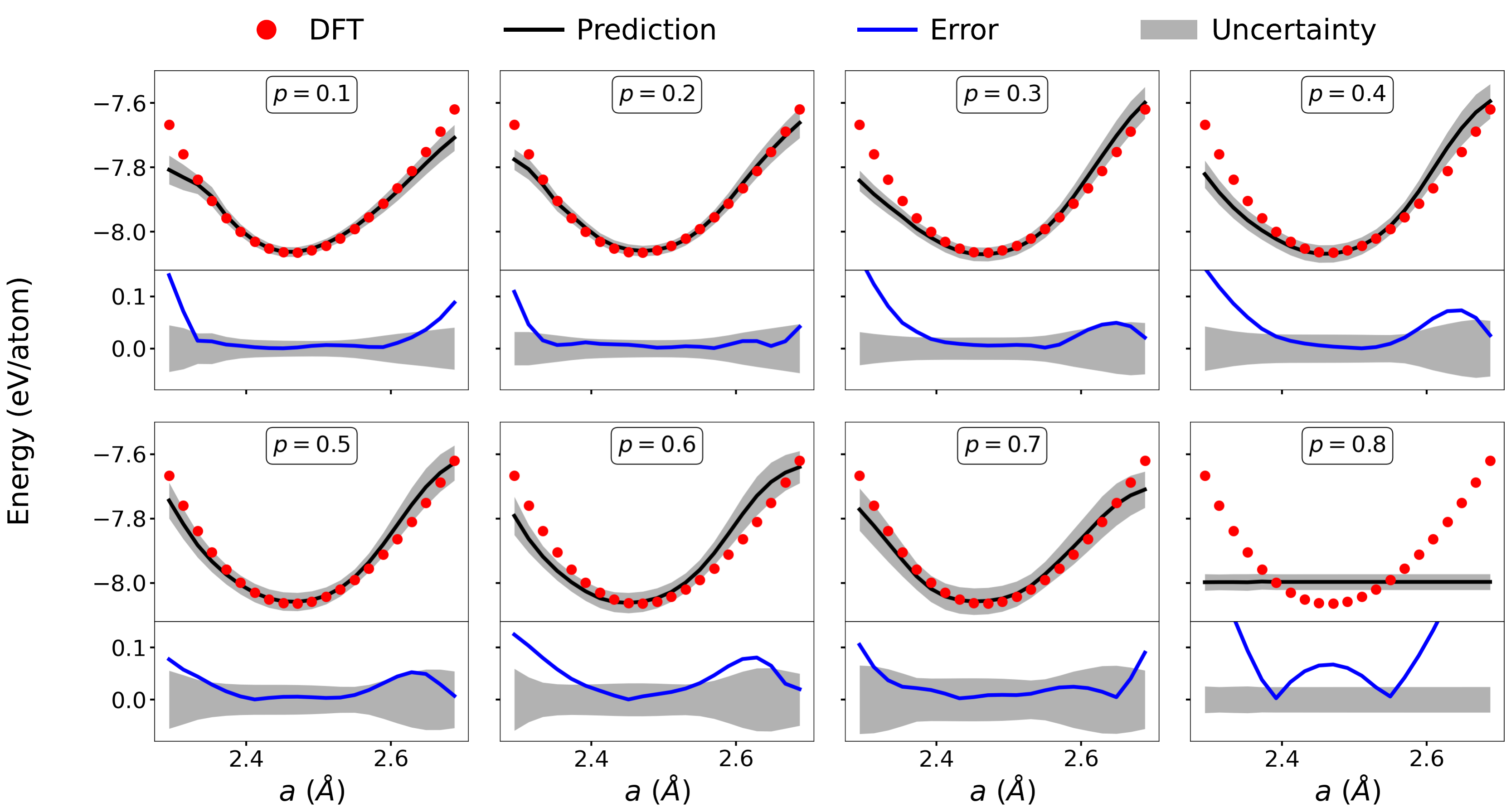}
    \caption[Graphene energy cold curve predictions and uncertainties across diffferent dropout ratios]{
	Energy cold curve predictions and uncertainties for graphene structure across several dropout ratios.
    }
    \label{fig:energy_latconst_graphene}
\end{figure*}

\begin{figure*}[!hbt]
    \centering
    \includegraphics[width=0.9\textwidth]{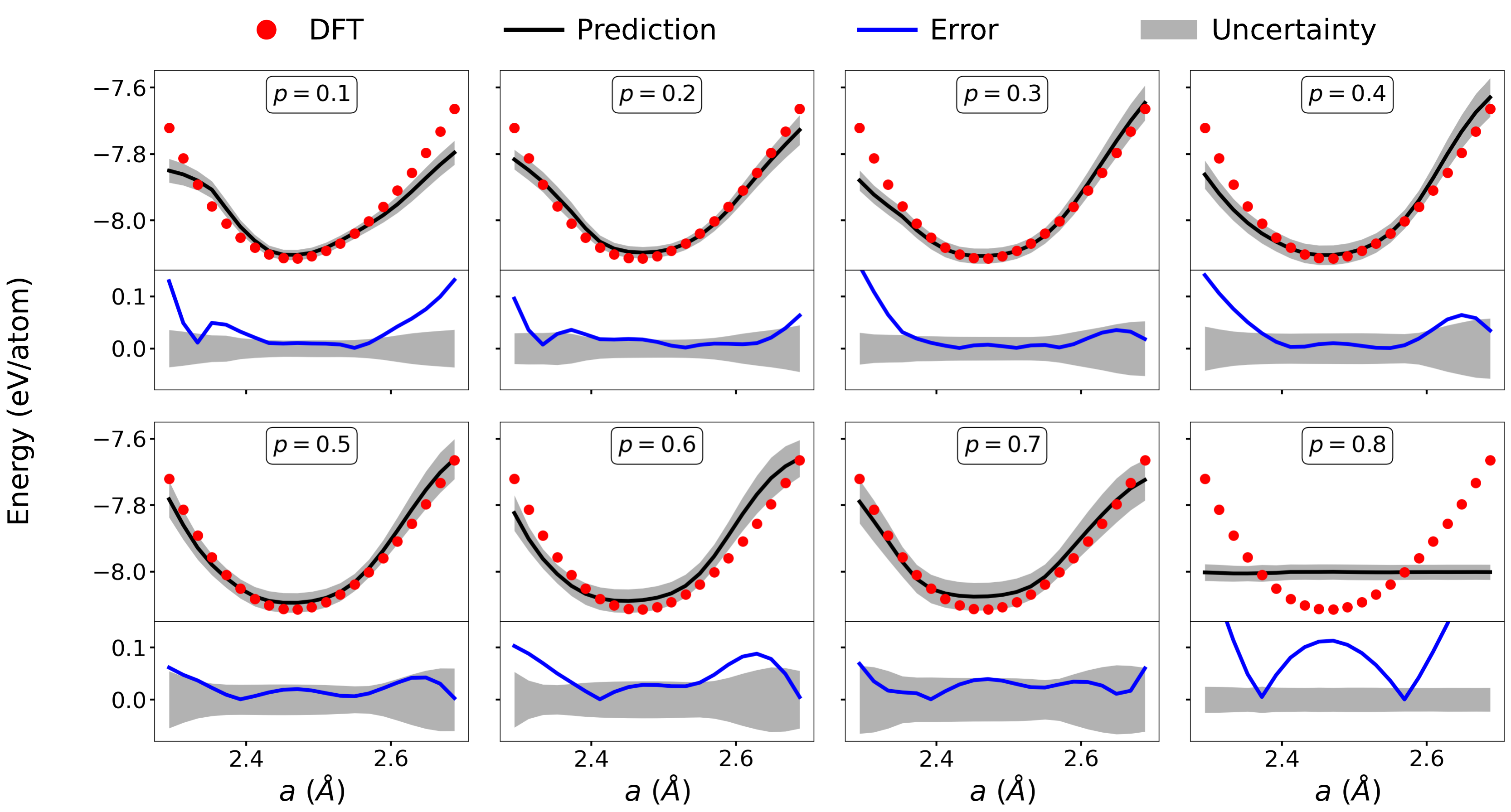}
    \caption[Graphite energy cold curve predictions and uncertainties across diffferent dropout ratios]{
	Energy cold curve predictions and uncertainties for graphite structure across several dropout ratios.
    }
    \label{fig:energy_latconst_graphite}
\end{figure*}

\clearpage
\subsection{Phonon dispersion}
\label{sec:phonon-dispersion}

In the following, we present the predicted phonon dispersion relations and their associated uncertainties for graphene, graphite, and diamond structures, obtained using different ensemble-based UQ models.
The prediction trends broadly mirror those observed in the energy cold curve analyses.
For graphene and graphite, the phonon spectra are generally well reproduced, although the predicted phonon energies are slightly underestimated in several modes, particularly the flexural optical mode near the $\Gamma$ point.
In contrast, the predictions for diamond are significantly less accurate.
The ensemble-averaged phonon dispersions deviate substantially from the DFT reference, especially in the optical branches, where the predicted frequencies are systematically underestimated.
The acoustic branches also show noticeable discrepancies, with the predicted curves appearing noisy, particularly near the $\Gamma$ point.
Although the uncertainty envelopes are wide in several regions---suggesting low model confidence---the DFT values frequently lie outside these bounds, indicating that the uncertainty estimates are overconfident and unreliable.
While some qualitative features of the phonon structure are preserved, the predictions fail to capture the correct vibrational behavior of the diamond lattice.

\begin{figure*}[!hbt]
    \centering
    \includegraphics[width=0.9\textwidth]{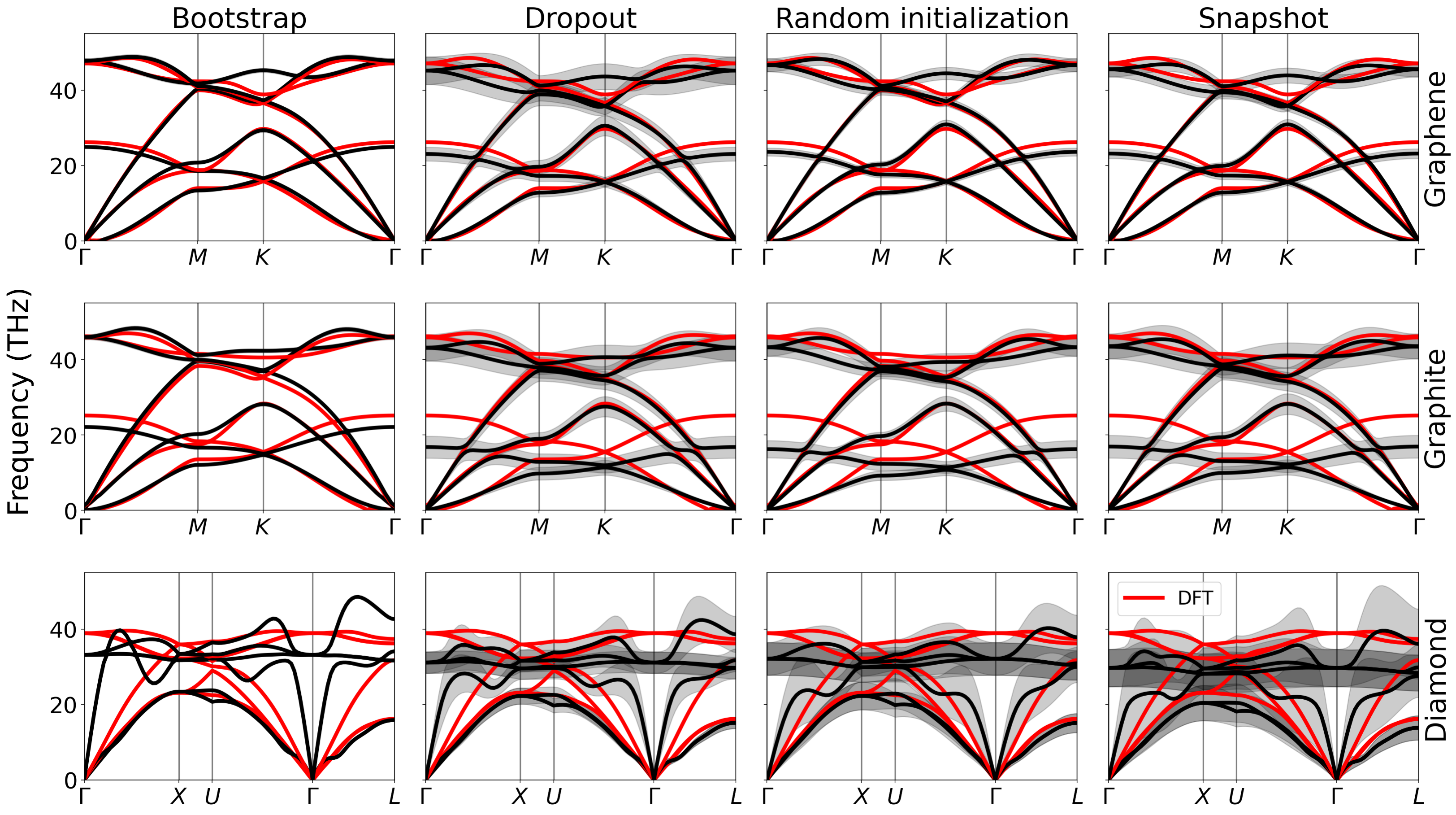}
    \caption[Phonon dispersion energy curves for different ensemble models]{
	Phonon dispersion energy curves for (top) graphene, (middle) graphite, and (bottom) diamond structures using different ensemble models.
	The ensemble-averaged predictions are shown as black curves, with the grey envelope representing the one-standard-deviation uncertainty.
	DFT ground truth values (red curves) are overlaid for comparison with the predicted values.
    }
    \label{fig:phonon_dispersion}
\end{figure*}

Comparisons of the phonon dispersion predictions and their associated uncertainties, obtained using progressively larger dropout ratios for graphene, graphite, and diamond structures, are shown below.
The phonon energies tend to be increasingly underestimated as the dropout ratio increases, indicating a degradation in the model’s learning capacity.
As in the energy cold curve predictions, this degradation can lead to unphysical results, particularly for high dropout values (e.g., $p \geq 0.8$).
Furthermore, although the associated uncertainties also increase with dropout ratio, the growth is insufficient to offset the rising errors, and the predictions remain overconfident.

\begin{figure*}[!hbt]
    \centering
    \includegraphics[width=0.9\textwidth]{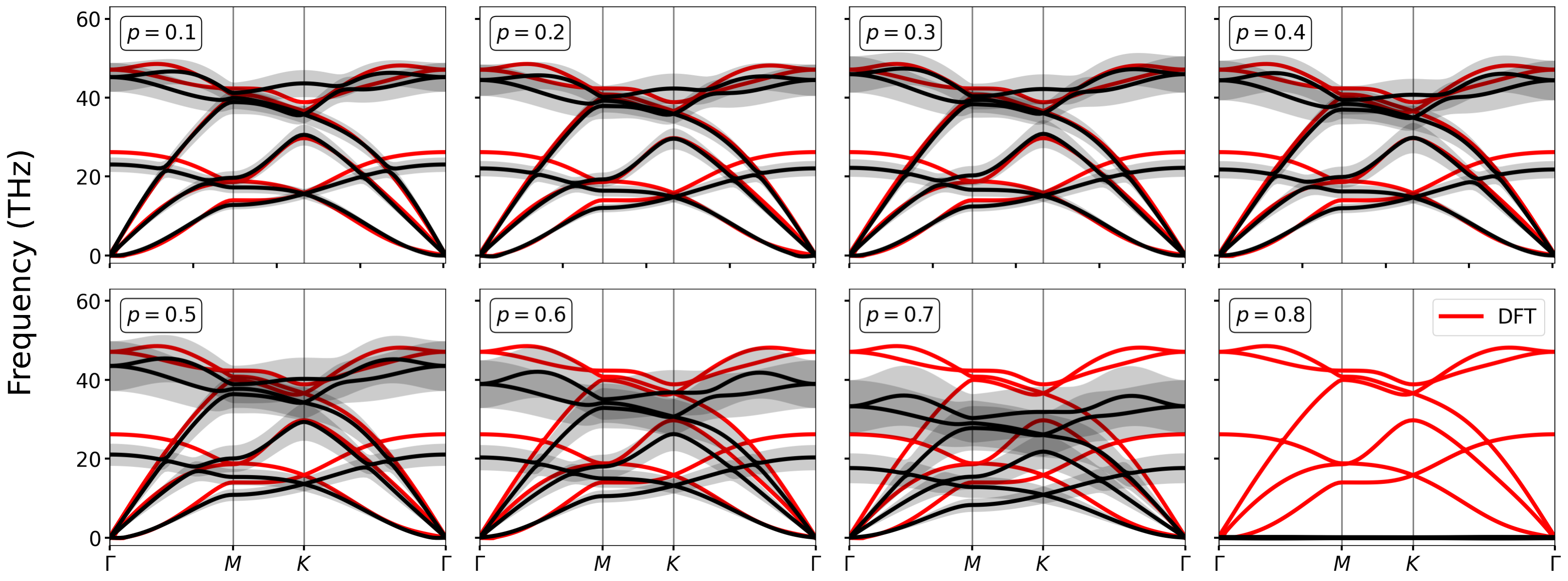}
    \caption[Graphene phonon dispersion energy predictions and uncertainties with across diffferent dropout ratios]{
	Phonon dispersion energy results for graphene structure with progressively increasing dropout ratio.
    }
    \label{fig:phonon_dropout_graphene}
\end{figure*}

\begin{figure*}[!hbt]
    \centering
    \includegraphics[width=0.9\textwidth]{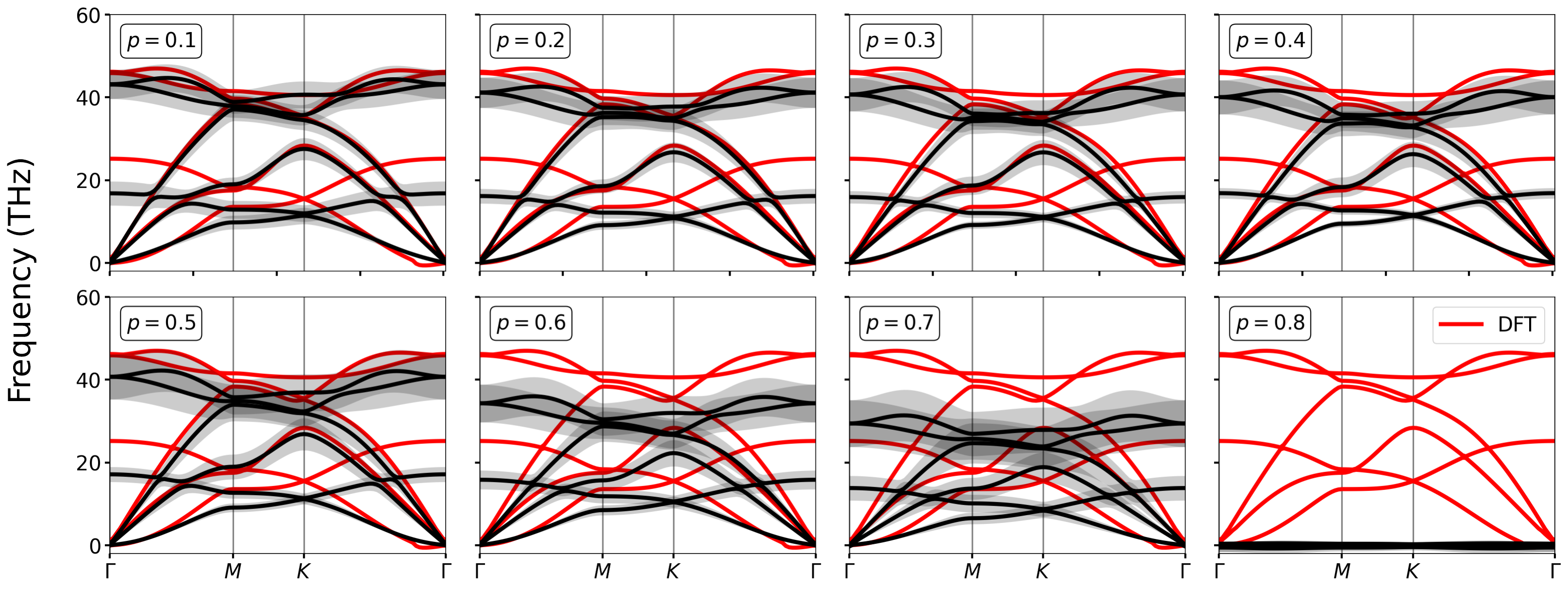}
    \caption[Graphite phonon dispersion energy predictions and uncertainties with across diffferent dropout ratios]{
	Phonon dispersion energy results for graphite structure with progressively increasing dropout ratio.
    }
    \label{fig:phonon_dropout_graphite}
\end{figure*}

\begin{figure*}[!hbt]
    \centering
    \includegraphics[width=0.9\textwidth]{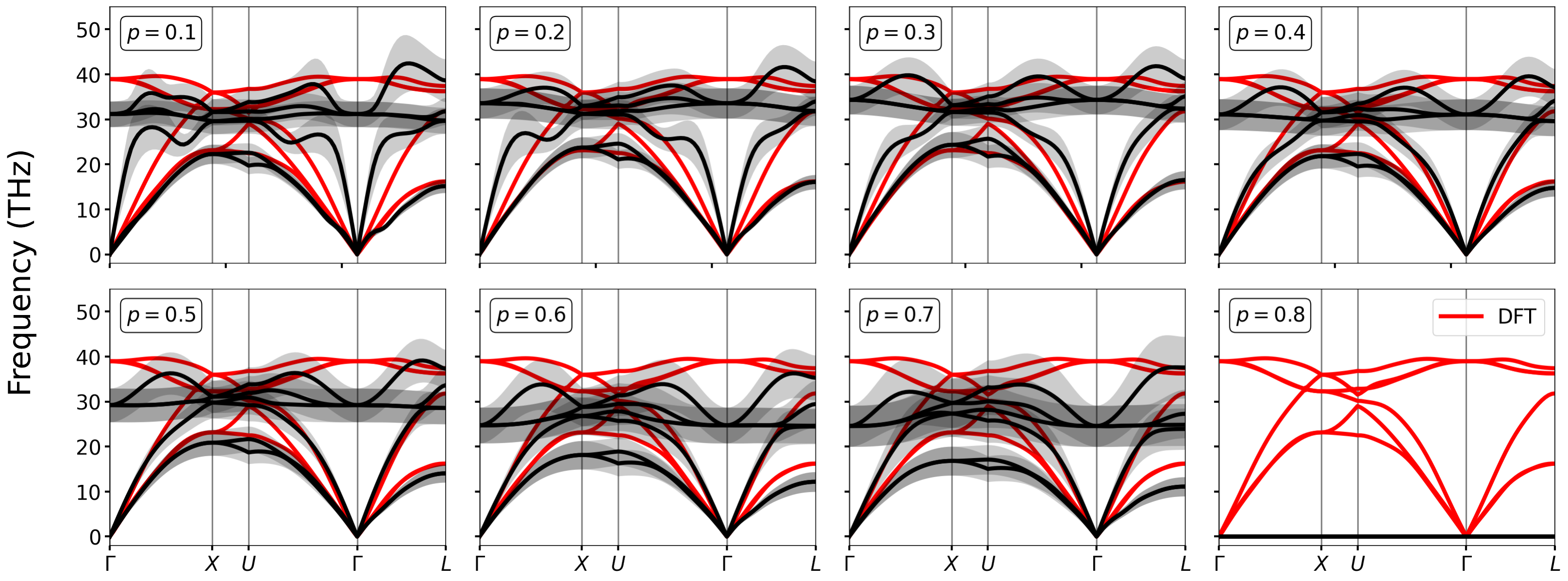}
    \caption[Diamond phonon dispersion energy predictions and uncertainties with across diffferent dropout ratios]{
	Phonon dispersion energy results for diamond structure with progressively increasing dropout ratio.
    }
    \label{fig:phonon_dropout_diamond}
\end{figure*}

\FloatBarrier

\bibliographystyle{plain}
\bibliography{refs,refs_zotero}